\documentclass[%
 reprint,
superscriptaddress,
 amsmath,amssymb,
 aps,
]{revtex4-1}

\usepackage{graphicx}
\usepackage{dcolumn}
\usepackage{bm}
\usepackage{braket}
\usepackage[title]{appendix}

\usepackage{xcolor}

\begin{document}

\preprint{APS/123-QED}

\title{A theoretical approach for electron dynamics and ultrafast spectroscopy (EDUS)}

\author{Giovanni Cistaro}
\affiliation{Departamento de Qu\'{i}mica, Universidad Aut\'{o}noma de Madrid, 28049 Madrid, Spain}
\author{Mikhail Malakhov}%
\affiliation{Departamento de Qu\'{i}mica, Universidad Aut\'{o}noma de Madrid, 28049 Madrid, Spain}
\author{Juan Jos\'e Esteve-Paredes}%
\affiliation{Departamento de F\'isica de la Materia Condensada, Universidad Aut\'{o}noma de Madrid, 28049 Madrid, Spain}
\author{Alejandro Jos\'e Ur\'ia-\'Alvarez}%
\affiliation{Departamento de F\'isica de la Materia Condensada, Universidad Aut\'{o}noma de Madrid, 28049 Madrid, Spain}
\author{Rui E. F. Silva}%
\affiliation{Instituto de Ciencia de Materiales de Madrid (ICMM), Consejo Superior de Investigaciones Científicas (CSIC),
Sor Juana Inés de la Cruz 3, 28049 Madrid, Spain}
\author{Fernando Mart\'in}
\affiliation{Departamento de Qu\'{i}mica, Universidad Aut\'{o}noma de Madrid, 28049 Madrid, Spain}
\affiliation{Instituto Madrile\~no de Estudios Avanzados en Nanociencia (IMDEA-Nanociencia), Cantoblanco, 28049 Madrid, Spain}
\affiliation{Condensed Matter Physics Center (IFIMAC), Universidad Autónoma de Madrid, 28049 Madrid, Spain}
\author{Juan Jos\'e Palacios}%
\affiliation{Departamento de F\'isica de la Materia Condensada, Universidad Aut\'{o}noma de Madrid, 28049 Madrid, Spain}
\affiliation{Condensed Matter Physics Center (IFIMAC), Universidad Autónoma de Madrid, 28049 Madrid, Spain}
\affiliation{Instituto Nicol\'as Cabrera, Universidad Aut\'{o}noma de Madrid, 28049 Madrid, Spain}
\author{Antonio Pic\'on}
\email{antonio.picon@uam.es, corresponding author}
\affiliation{Departamento de Qu\'{i}mica, Universidad Aut\'{o}noma de Madrid, 28049 Madrid, Spain}

\date{\today}

\begin{abstract}
In this manuscript we present a theoretical framework and its numerical implementation to simulate the out-of-equilibrium electron dynamics induced by the interaction of ultrashort laser pulses in condensed-matter systems. Our approach is based on evolving in real-time the density matrix of the system in reciprocal space. It considers excitonic and non-perturbative light-matter interactions. We show some relevant examples that illustrate the efficiency and flexibility of the approach to describe realistic ultrafast spectroscopy experiments. Our approach is suitable for modeling the promising and emerging ultrafast studies at the attosecond time scale that aim at capturing the electron dynamics and the dynamical electron-electron correlations via X-ray absorption spectroscopy.
\end{abstract}

\maketitle


\section{\label{sec:intro}Introduction}

Optical manipulation is the fastest technique to control and switch properties in a material. The advent of ultrashort laser pulses enable to drive the system out-of-equilibrium and reach novel quantum phases with properties beyond the ones at  equilibrium \cite{Basov2017}. Modifications of the topological phase \cite{Oka2009,Inoue2010,Sie2015,McIver2020,JimenezGalan2020}, control of the valley pseudo-spin via optical resonant excitation \cite{Xiao2010,Xu2014,Langer2018}, coherent light-driven currents \cite{Higuchi2017,Reimann2018,Heide2018}, and light-induced insulator-to-conductor transitions \cite{Schiffrin2013,Schultze2013,Mashiko2016} are some promising applications within the out-of-equilibrium phenomena.

Significant progress has been achieved in recent years to implement time-resolved experiments for tracking and reading out  transient state dynamics within the non-equilibrium system. Remarkably, it is nowadays possible to follow the electron dynamics in its natural time scale, i.e. in the attosecond timescale (10$^{-18}$ s), and well before the lattice starts to respond to the external field. Attosecond transient absorption spectroscopy (ATAS) is a promising technique for tracking electron dynamics based on a pump-probe scheme, typically the pump being a IR/mid-IR few-femtosecond pulse and the probe being an attosecond XUV/soft-x-ray attosecond pulse \cite{Leone2016}. The power of ATAS is that it combines attosecond temporal resolution with high energy resolution, much higher than that provided by photoelectron spectroscopy. ATAS has been successfully applied in different bulk and thin materials, from insulators to semimetals, in order to investigate carrier dynamics, phononic effects, and excitonic interactions \cite{Schultze2014,Lucchini2016,Zuerch2017,Moulet2017,Schlaepfer2018,Volkov2019,Lucchini2021,Buades2021}. 

In this context, there is a natural need to simulate the transient state dynamics of condensed matter systems under laser excitation, both for understanding the underlying mechanisms of out-of-equilibrium properties and for correlating the microscopic electron dynamics with the macroscopic measurements, observables such as current and absorption. The modeling of time-resolved experiments demands an additional complexity. In those, the probe pulse provides information of the out-of-equilibrium system for a specific time delay between the pump and probe pulse. However, the probe pulse must be included in the model, as it contributes to the dynamics and is directly linked to the measured of observables at particular time delays. Furthermore, the typical intensities of IR/mid-IR ultrashort pulses enable nonlinear interactions that must be accounted for in the model. Also, pump-probe schemes can be viewed as nonlinear schemes, as they require the absorption, at least, of two photons at two different times. Last but not least, in all non-metallic two-dimensional materials known to date the optical response is dominated by excitonic effects. This is, to a good extent, due to a suppressed screening of interactions in low dimensions, which facilitates the binding between electrons and holes \cite{Wang2018}. Excitons can be considered as quasi-particles composed of an electron-hole pair bound via Coulomb interaction. Hence, electron dynamics simulations must be able to describe the formation of excitons in order to properly describe the light-matter interaction. In this context, we aim at covering all these demands and we present in this manuscript a theoretical approach that allows us to simulate electron dynamics in realistic condensed-matter systems driven out of equilibrium as well as to model ultrafast/time-resolved spectroscopy experiments. 

Density-functional theory (DFT) is the workhorse of computational modeling for materials at equilibrium. However, out-of-equilibrium dynamics is beyond the scope of DFT and there are three main alternatives. The first one is time-dependent DFT (TDDFT) \cite{Runge1984, Marques2012,Yabana2012}, which consists in solving a time-dependent Kohn-Sham equation. There are several TDDFT codes implemented in real-time and in real-space ideal for condensed-matter systems interacting with laser pulses, see for example Refs. \cite{octopus,salmon}. In order to account for excitonic interactions in TDDFT, a long-range nonlocal exchange functional is needed and the numerical implementation thus implies a high computational cost \cite{Abramson2015}. The second one is based on many-body perturbation theory (MBPT). Starting from the Kohn-Sham DFT electronic structure, the well-known Bethe-Salpeter equation (BSE) is solved and the energy and wavefunctions of excitons are obtained, the latter expressed as a superposition of single particle excitations \cite{Rohlfing1998,Albrecht1998}. BSE provides accurate energies and it is ideal for spectroscopy calculations. However, BSE is not a time-domain framework, it cannot describe real-time non-equilibrium dynamics and ultrafast spectroscopy experiments. In recent years there have been several theoretical approaches to extend BSE in the time domain \cite{Attaccalite2011,Ridolfi2020,Sangalli2021}. This consists in resolving the Kadanoff-Baym equations based on the nonequilibrium Green's function theory \cite{Rink1988}. 

The third method is similar to solving the Kadanoff-Baym equations for the Green function, but starting instead from a second quantization formalism and evolving the reduced density matrix. Within this formalism the well-known semiconductor Bloch equations can be derived \cite{Haug2004,Kira2006}, which evolve the density matrix of the system in reciprocal space. Our approach is based on this theoretical framework. We show how the equations of motion for the density matrix can be efficiently implemented. We model relevant physical scenarios that illustrate the flexibility of our approach through either simple tight-binding (TB) models or within the Kohn-Sham DFT scheme, both with localized orbitals or Wannier basis, the description of excitonic effects in optical absorption spectra, and the feasibility to model ATAS using attosecond X-ray pulses both in 2D and 3D materials.

\section{\label{sec:theory}Theoretical framework}

In this section we present the main theory for the evolution of the density matrix of a periodic system interacting with laser pulses. We detail the main approximations that are used in the numerical implementation.

\subsection{The density matrix}
The many body state of a periodic system can be represented in the second quantization formalism as
\begin{equation}
\vert \psi \rangle = \prod_{n{\bf k}} \hat{c}^\dagger_{n\bf k} \vert 0 \rangle
\end{equation}
in which $\vert 0 \rangle$ represents the vaccum, and the canonical operator $\hat{c}^\dagger_{n\bf k}$ applying on the vacuum creates an electron in the state $\vert n{\bf k} \rangle$. The quantum number ${\bf k}$ refers to the quasi-momentum, while $n$ refers to the band (energy) level and spin. Typically, in the equilibrium, the state is populated up to a certain energy, the so-called Fermi energy or level. For a nonzero temperature, the equilibrium state should be represented by a statistical incoherent ensemble. 

When a laser pulse interacts with the system at the equilibrium, then the many-body state will evolve in time $\vert \psi (t) \rangle$. In order to describe the system evolution that is driven out of the equilibrium, one can use the reduced, one particle density matrix 
\begin{equation} \label{DensityMatrix}
\rho_{nm}(\textbf{k},t) \equiv \langle \hat{c}^\dagger_{m{\bf k}} \hat{c}_{n{\bf k}} \rangle = {\rm Tr} [\hat{\varrho}(t)  \hat{c}^\dagger_{m{\bf k}} \hat{c}_{n{\bf k}} ]
\end{equation}
where $\hat{\varrho}(t) \equiv \vert \psi (t) \rangle \langle \psi(t) \vert$ is the evolving density operator, which can be easily generalized for a statistical incoherent ensemble. Note that in general one could use $\langle \hat{c}^\dagger_{m{\bf k}} \hat{c}_{n{\bf k}'} \rangle$, but we will show in the following that the form given by Eq. (\ref{DensityMatrix}) is sufficient to capture the evolution of the system.

\subsection{The many-body Hamiltonian}

The total Hamiltonian of a periodic system is expressed as
\begin{eqnarray}
\hat{H}_{sys}(t) = \hat{H}_{0} + \hat{H}_{e-e} + \hat{H}_{I}(t)
\end{eqnarray}
in which $\hat{H}_{0}$ contains the non-interacting terms of all electrons, $\hat{H}_{e-e}$ accounts for the electron-electron interactions, and $\hat{H}_{I}$ is the laser-matter interaction. The latter term depends on the external electric field of the laser pulse and is then time-dependent. Other contributions that could arise from the lattice motion, such as phonon interactions, are neglected. This is justified because we aim at exploring very short time scales of the out-of-equilibrium system in the range of few femtoseconds, in which the electron motion will be the relevant contribution. In particular, the different Hamiltonians are
\begin{eqnarray}
\hat{H}_{0} &=& \sum_{n, \mathbf{k}}\varepsilon^0_{n, \mathbf{k}}\hat{c}^{\dagger}_{n, \mathbf{k}}\hat{c}_{n, \mathbf{k}} \\
\hat{H}_{e-e} &=& \frac{1}{2}  \sum_{n,m, \mathbf{k}, \mathbf{k}', \mathbf{q}} \!\!\!\! U_{nm,{\bf k k' q}} \, \hat{c}^{\dagger}_{n, \mathbf{k+q}}\hat{c}^{\dagger}_{m, \mathbf{k'-q}} \hat{c}_{m, \mathbf{k'}}\hat{c}_{n, \mathbf{k}} \\
\hat{H}_{I} &=& |e| \sum_{n, m,{\bf k}} c_{n {\bf k}}^{\dagger}\, \boldsymbol{\varepsilon}(t) \cdot\left[i \delta_{n m} \nabla_{{\bf k}}+\boldsymbol{\xi}_{n m}({\bf k})\right] c_{m {\bf k}}
\end{eqnarray}
where $\varepsilon^0_{n, \mathbf{k}}$ is the energy dispersion of the band $n$, $\boldsymbol{\xi}_{n m}$ is the Berry connection, $\boldsymbol{\varepsilon}(t)$ is the electric field of the laser pulse, and $U_{nm,{\bf k k' q}}$ is the Coulomb interaction between two particles defined as $U_{nm,{\bf k k' q}} \equiv W_{n{\bf k+q} \, m{\bf k'-q},m{\bf k}'n{\bf k}}$, where
\begin{eqnarray} \label{CoulombW}
W_{m\textbf{k}l\textbf{k}',j\textbf{k}'_1 n\textbf{k}_1} =\int\int_{ \Omega} d^3 r d^3 r' \times \hspace{2cm} \nonumber\\ \psi^*_{m\textbf{k}}(\textbf{r})\psi^*_{l\textbf{k}'}(\textbf{r}')V(\textbf{r}-\textbf{r}') \psi_{j\textbf{k}'_1}(\textbf{r}')\psi_{n\textbf{k}_1}(\textbf{r})
\end{eqnarray}
being $V(\textbf{r}-\textbf{r}')$ the Coulomb energy between two particles and $\psi_{n\textbf{k}} ({\bf r}) \equiv \langle {\bf r} \vert n {\bf k} \rangle $.

The light-matter interaction term is written in the so-called length gauge $\hat{H}_I (t) = |e|\boldsymbol \varepsilon(t) \cdot \hat{\textbf{r}}$, where $\hat{\bf r}$ represents the position operator of the electrons in the system. This particular form is only valid in the dipole approximation, when the size of the quantum system is smaller than the wavelength of the external electric field. For a unit cell, whose typical size is of few nanometers, this approximation is well-justified for optical and IR wavelengths. However, it may be compromised for wavelengths lower than 1 nm, corresponding to photon energies larger than 1.24 keV (x-ray regime). The position operator can be expressed as 
\begin{eqnarray}
\hat{\bf r} = \sum_{{\bf k}',{\bf k}}\sum_{n, m} \langle n,{\bf k}' \vert \hat{\bf r}_1 \vert m,{\bf k} \rangle c_{n {\bf k}'}^{\dagger} c_{m {\bf k}}
\end{eqnarray}
being $\vert n, {\bf k} \rangle$ is one electron state defined as $\vert n, {\bf k} \rangle \equiv \hat{c}_{n,{\bf k}} \vert 0 \rangle$ and $\hat{\bf r}_1$ is the position operator acting on one particle of the indistinguishable system. In general, for any basis satisfying the Bloch theorem, i.e. with the form $\langle {\bf r} \vert n, {\bf k} \rangle = \psi_{n,\bf k}(\textbf{r}) = e^{i\textbf{k}\cdot \textbf{r}} u_{n,\textbf{k}}(\textbf{r})$, where $u_{n,\textbf{k}}(\textbf{r})$ is a periodic function, the transition element is 
\begin{eqnarray}
\langle n,{\bf k}' \vert \hat{\bf r}_1 \vert m,{\bf k} \rangle = -i \delta_{nm}\boldsymbol{\nabla}_{\textbf{k}} \delta({\bf k} - {\bf{k}}') + {\pmb{\xi}}_{nm}({\bf k})\delta({\bf k} - {\bf{k}}') \nonumber \\
\end{eqnarray}
in which the Berry connection is written as:
\begin{equation}
\pmb{\xi}_{nm}({\bf k}) = i {1\over\Omega_{uc}}\int_{\Omega_{uc}} d^3 r \; u^*_{n{\bf k}}(\textbf{r}){\boldsymbol\nabla}_{\bf k} u_{m\bf k}(\textbf{r})
\end{equation}
The spatial integration is performed only over a unit cell $\Omega_{uc}$. This transition element gives rise to the used light-matter interaction Hamiltonian in the length gauge.

The numerical implementation of the $\hat{H}_{e-e}$ term, which also accounts for exciton-exciton interactions, requires a high computational effort. In order to reduce this effort, one can make the mean-field approximation that transforms this term in an effective single-particle operator
\begin{eqnarray}
\hat{H}_{e-e} &\approx& - \sum_{n,m, \mathbf{k}, \mathbf{k}', \mathbf{q}} U_{nm,{\bf k k' q}} \langle \hat{c}^{\dagger}_{n, \mathbf{k+q}} \hat{c}_{m, \mathbf{k'}} \rangle \hat{c}^{\dagger}_{m, \mathbf{k'-q}} \hat{c}_{n, \mathbf{k}} \nonumber  \\ 
\end{eqnarray}
In this approximation one relies on the fact that the contribution $\hat{c}^{\dagger}_n \hat{c}_{m}$ in the Coulomb interaction is well-described by the average $\langle\hat{c}^{\dagger}_n \hat{c}_{m}\rangle$, see Ref. \cite{Bruus}. We assume excitations ${\bf q}={\bf k}'-{\bf k}$ in which excitons do not carry momentum, i.e. $\langle\hat{c}^{\dagger}_{m, \mathbf{k}} \hat{c}_{n, \mathbf{k'}}\rangle$ = 0 if $\mathbf{k} \neq \mathbf{k}'$, i.e.
\begin{eqnarray} \label{HeeMF}
\hat{H}_{e-e} &\approx& - \sum_{n,m, \mathbf{k}, \mathbf{k}'} W_{n\textbf{k}'m\textbf{k},m\textbf{k}' n\textbf{k}} \langle \hat{c}^{\dagger}_{n, \mathbf{k'}} \hat{c}_{m, \mathbf{k'}} \rangle \hat{c}^{\dagger}_{m, \mathbf{k}} \hat{c}_{n, \mathbf{k}} \nonumber \\
&=& - \sum_{n,m, \mathbf{k}, \mathbf{k}'} W_{n\textbf{k}'m\textbf{k},m\textbf{k}' n\textbf{k}} \, \rho_{mn}(\textbf{k}',t) \hat{c}^{\dagger}_{m, \mathbf{k}} \; \hat{c}_{n, \mathbf{k}} \nonumber \\
\end{eqnarray}
The average term $\langle\hat{c}^{\dagger}_n \hat{c}_{m}\rangle$ corresponds exactly with the density matrix $\rho_{mn}(\textbf{k},t)$ that we propagate in time. Hence, the Hamiltonian (\ref{HeeMF}) requires to be computed at each time step.

\subsection{Equations of motion}

By using the von Neumman equation for the evolution of the density operator, one obtain the equations of motion (EOM) for the one particle density matrix (\ref{DensityMatrix}) via $ i \hbar \, \partial \rho_{n m}/\partial t = {\rm Tr} ([{\hat{H}_{sys}}(t),\hat{\varrho}(t)] \hat{c}^\dagger_{m} \hat{c}_{n})$. In particular, the equations of motion can be reduced to

\begin{eqnarray} \label{EOM}
 i\hbar \frac{\partial \rho_{nm}(\textbf{k},t) }{\partial t} =  i\hbar \frac{\partial \rho_{nm}(\textbf{k},t) }{\partial t} \vert_{deph} \hspace{3cm} \nonumber \\
 + \left[H_0 (\textbf{k}) + H_{e-e} (\textbf{k}) + |e| {\boldsymbol \varepsilon}(t)\cdot \boldsymbol{\xi}(\textbf{k}), \, \rho(\textbf{k},t)\right]_{nm} \nonumber \\
 +i |e| \boldsymbol{\varepsilon}(t) \cdot \nabla_\textbf{k}\, \rho_{nm}(\textbf{k},t) \nonumber \\
\end{eqnarray}
where the matrix form is used, i.e. 
\begin{eqnarray*}
H_0 (\textbf{k})\vert_{nm} &=& \varepsilon^0_{n, \mathbf{k}} \delta_{nm} \\
H_{e-e} (\textbf{k})\vert_{nm} &=& - \sum_{\mathbf{k}'} W_{n\textbf{k}'m\textbf{k},m\textbf{k}' n\textbf{k}} \; \rho_{nm}({\bf k}',t)  \\
\boldsymbol{\xi}(\textbf{k}) \vert_{nm} &=& \boldsymbol{\xi}_{nm} ({\bf k})
\end{eqnarray*}
Note that the two light-matter interaction terms, which depend on the external electric laser field $\boldsymbol{\varepsilon}(t)$, are quite different. While the one with the Berry connections may induce interband and intraband transitions, the one with the gradient of the quasi-momentum ${\bf k}$ is purely intraband. The latter, which is related to the semiclassical electron propagation, is important for IR and THz fields and cannot be neglected for typical intensities between $10^9$ and $10^{12}$ W/cm$^2$. Note also that electron-electron interactions make the EOM to be nonlinear with respect to the density matrix. All electron correlations beyond the mean field approximation are included in the term $i\hbar {\partial \rho_{nm}(\textbf{k},t) }/{\partial t} \vert_{deph}$. Here, it is customary to include relaxation and dephasing effects arising from electron-electron interactions and electron-phonon couplings. 

\subsection{The Bloch gauge}
In the previous section, the equations of motion were introduced in the eigenstate basis, i.e. a basis in which the non-interacting Hamiltonian ${\hat{H}_0}$ is diagonal and satisfies ${\hat{H}_0} \vert n,{\bf k} \rangle = {\hat{H}_0} \hat{c}_{n,{\bf k}} \vert 0 \rangle = \varepsilon_{n,\bf k}^0 \hat{c}_{n,{\bf k}} \vert 0 \rangle $. However, it may be convenient to simulate the dynamics in another basis in order to reduce the computational cost of the equations of motion given by Eq. (\ref{EOM}).  In general, one can write a Bloch basis set as a linear combination of well-localized functions $\vert \alpha, \textbf{R} \rangle$:

\begin{equation}\label{Bloch_basis}
\vert \alpha, \textbf{k} \rangle = \frac{1}{\sqrt{N}} \sum_{\textbf{R}} e^{i\textbf{k}\cdot \textbf{R}} \vert \alpha, \textbf{R} \rangle
\end{equation}

Ideally, $\vert \alpha, \textbf{R} \rangle$ are functions that in real space decay exponentially (for instance, Gaussian orbitals or Wannier functions).
This basis set satisfies Bloch's theorem, even if it does not diagonalize the Hamiltonian $\hat{H}_0$. The sum of ${\bf R}$ goes over all unit cells of the system, and $N$ represents the total number of units cells. In this basis, note that the ${\bf k}$ dependence is in the imaginary exponential function of equation (\ref{Bloch_basis}). Hence, the states are smooth functions of ${\bf k}$. This has a clear advantage to simulate the dynamics, in which a less fine numerical ${\bf k}$-grid would be needed, in contrast with the eigenstate basis, which present less smooth states and even singular points in the reciprocal space.  In general, we can always express the eigenstates $\vert n, {\bf k} \rangle$ of our non-interacting Hamiltonian with respect to our Bloch basis, i.e.
\begin{equation} \label{unitary}
\vert n, {\bf k} \rangle = \sum_{\alpha} U_{\alpha,n} ({\bf k}) \vert \alpha, {\bf k} \rangle
\end{equation}
or similarly $\hat{c}_{n,{\bf k}} = \sum_\alpha U_{\alpha,n} \hat{c}_{\alpha,{\bf k}}$, in which $\hat{c}_{\alpha,{\bf k}}$ is an operator that creates an electron in our Bloch basis. Our change of basis is represented by a unitary matrix $U\vert_{\alpha,n} = U_{\alpha,n}$ which may depend on the quasi momentum. Vice-versa, one can write the Bloch basis in the eigenstate basis:
\begin{equation}
\vert \alpha, {\bf k} \rangle = \sum_{n} U^\dagger_{n,\alpha} ({\bf k}) \vert n, {\bf k} \rangle
\end{equation}

Now, one can reformulate the equations of motion in the Bloch basis, defining the density matrix in this basis as $\rho_{\alpha\beta}^{(B)}(\textbf{k},t) \equiv \langle \hat{c}^\dagger_{\beta{\bf k}} \hat{c}_{\alpha{\bf k}} \rangle $

\begin{eqnarray} \label{EOM_Bloch}
 i\hbar \frac{\partial \rho_{\alpha\beta}^{(B)}(\textbf{k},t) }{\partial t} =  i\hbar \frac{\partial \rho_{\alpha\beta}^{(B)}(\textbf{k},t) }{\partial t} \vert_{deph} \hspace{3cm} \nonumber \\
 \left[H_0^{(B)} (\textbf{k}) + H_{e-e}^{(B)} (\textbf{k}) + |e| {\boldsymbol \varepsilon}(t)\cdot \boldsymbol{\xi}^{(B)}(\textbf{k}), \, \rho^{(B)}(\textbf{k},t)\right]_{\alpha\beta} \nonumber \\
 +i |e| \boldsymbol{\varepsilon}(t) \cdot \nabla_\textbf{k}\, \rho_{\alpha\beta}^{(B)}(\textbf{k},t) \nonumber \\
\end{eqnarray}

Now the different Hamiltonians are written in the Bloch basis and they are connected with the eigenstate basis by 

\begin{eqnarray*}
H_0 ({\bf k}) &=& U^\dagger ({\bf k}) H_0^{(B)}({\bf k}) U({\bf k}) \\
H_{e-e} ({\bf k}) &=& U^\dagger ({\bf k}) H_{e-e}^{(B)}({\bf k}) U({\bf k})\\
\boldsymbol{\xi} ({\bf k}) &=& U^\dagger ({\bf k}) \boldsymbol{\xi}^{(B)}({\bf k}) U({\bf k}) + i U^\dagger ({\bf k}) \nabla_\textbf{k} U({\bf k})
\end{eqnarray*}

Note that the electron-electron transition elements $H_{e-e}$ are computed at each time step through the density matrix at the eigenstate basis. In the Bloch basis, one may compute $H_{e-e}$ by which first a basis change of the density matrix $\rho ({\bf k},t) = U^\dagger ({\bf k}) \rho^{(B)}({\bf k},t) U({\bf k})$ and then a second basis change of the electron-electron Hamiltonian. 

A gauge transformation can be considered as a unitary transformation such that $U({\bf k})\vert_{nm} = \delta_{nm} e^{-i \varphi_{m}({\bf k})}$. By using the previous formalism, it is easy to check that the EOM are preserved under a gauge transformation.

\subsection{The x-ray interactions}

In time-resolved experiments, the system is brought out of equilibrium by a laser pulse, the so-called pump pulse, and the dynamics are studied by a second laser pulse, the so-called probe pulse. In order to calculate the observables for the probe pulse, one needs to calculate the transient electron dynamics induced by both pulses. However, most of the times the probe pulse has a photon energy (frequency) much higher than the pump pulse. In attosecond science, typically the pump frequency is in the IR/mid-IR range, while the probe frequency is in the XUV/soft-X-ray range. Simulating the electron dynamics of both pulses is a challenging task, because one needs a fine resolution in time in order to properly describe the probe pulse effects, while the pump pulse forces a long integration in time. In order to reduce this effort, one may perform the rotating-wave approximation (RWA) on the probe interactions. The external electric field is split in $\boldsymbol{\varepsilon}(t) = \boldsymbol{\varepsilon}_{p}(t) + \boldsymbol{\varepsilon}_{x}(t)$, where $\boldsymbol{\varepsilon}_{p}(t)$ and $\boldsymbol{\varepsilon}_{x}(t)$ correspond to the pump and probe electric field, respectively. The probe pulse is described as ${\boldsymbol \varepsilon}_x(t) = {\bf g}_x(t) \cos{\omega_x t}$, where ${\bf g}_x(t)$ is a slowly-variant envelope function and $\omega_x$ is the central frequency. The probe interactions involve a core band and a valence/conduction band, and the corresponding off-diagonal term of the density matrix oscillates then very rapidly, with frequencies of the order of the central frequency $\omega_x$. Writing the off-diagonal term of the density matrix as $\rho_{nn_c} = e^{-i\omega_x t} {\tilde \rho}_{nn_c}$, where $n_c$ and $n$ labels refer to the core band and a valence/conduction band respectively, enables to separate the fast from the slowly-variant changes in time. Including this transformation in the EOM, Eq. (\ref{EOM}), and performing the RWA, i.e. neglecting the fast terms in time, one obtains \cite{Picon2019}

\begin{eqnarray} \label{EOM_RWA}
 i\hbar \frac{\partial \rho_{nm}^{(R)}(\textbf{k},t) }{\partial t} =  i\hbar \frac{\partial \rho_{nm}^{(R)}(\textbf{k},t) }{\partial t} \vert_{deph} + S_{nm}\, \rho_{nm}^{(R)}(\textbf{k},t)  \hspace{0.5cm} \nonumber \\
 +\left[H_0^{(R)} (\textbf{k}) + H_{e-e}^{(R)} (\textbf{k}) + |e| {\boldsymbol \varepsilon}^{(R)}(t)\cdot \boldsymbol{\xi}(\textbf{k}), \, \rho^{(R)}(\textbf{k},t)\right]_{nm} \nonumber \\
 +i |e| \boldsymbol{\varepsilon}(t) \cdot \nabla_\textbf{k}\, \rho_{nm}^{(R)}(\textbf{k},t) \nonumber \\
\end{eqnarray}
where the off-diagonal terms involving a core band $n_c$ and a valence/conduction band $n$ are cast as
\begin{eqnarray*}
\rho_{nn_c}^{(R)} &=&  {\tilde \rho}_{nn_c} \\
S_{nn_c} &=& -\hbar \omega_x \\
H_{e-e}^{(R)}\vert_{nn_c} &=& - \sum_{\mathbf{k}'} W_{n\textbf{k}'m\textbf{k},m\textbf{k}' n\textbf{k}} \, \tilde{\rho}_{nn_c}({\bf k}',t) \\
|e| {\boldsymbol \varepsilon}^{(R)}(t)\cdot \boldsymbol{\xi}(\textbf{k}) \vert_{nn_c} &=& |e| {{\bf g}_x(t)\over 2}\cdot \boldsymbol{\xi}(\textbf{k}) \vert_{nn_c}
\end{eqnarray*}
otherwise
\begin{eqnarray*}
\rho_{nm}^{(R)} &=&  { \rho}_{nm} \\
S_{nm} &=& 0 \\
H_{e-e}^{(R)}\vert_{nm} &=& {H}_{e-e} (\textbf{k})\vert_{nm} \\
|e| {\boldsymbol \varepsilon}^{(R)}(t)\cdot \boldsymbol{\xi}(\textbf{k}) \vert_{nm} &=& |e| {\boldsymbol \varepsilon}_p(t)\cdot \boldsymbol{\xi}(\textbf{k}) \vert_{nm}
\end{eqnarray*}

Note that also the excitonic interaction accounts for the slow-variant terms in the off-diagonal terms involving core electrons.


\subsection{Relaxation effects}
All electron correlations beyond the mean field approximation are included in the term $i\hbar {\partial \rho_{nm}(\textbf{k},t) }/{\partial t} \vert_{deph}$. This term includes electron-electron interactions that are not accounted for in the mean-field approximation, but also electron-phonon scattering effects. These terms are challenging to compute and some approximations are required to reduce their computational effort, but still they describe the main underlying physics. 

The decay of a core hole state is very fast in the order of hundreds of attoseconds to few femtoseconds, mainly triggered by Auger and fluorescence transitions. Ultrafast experiments with attosecond pulses are able to reach the soft-x-ray regime. At this regime it is possible to access the relevant K-edge absorption lines of carbon, nitrogen, and oxygen. For light elements below neon, Auger transitions are the dominant ones in the core-hole decay. Auger transitions are described by a four-operator Coulomb interaction, in which an electron from a valence shell occupies the core hole, and the energy release from this transition is transferred to another valence electron that will be promoted into the free continuum. Typically, those transitions are interpreted as a state coupled to a continuum, which plays the role of a quantum bath. Within the so-called Markov or local approximation \cite{Picon2017}, the effects on the core hole state can be reduced to a relaxation rate, i.e. $i\hbar {\partial \rho_{nm}(\textbf{k},t) }/{\partial t} \vert_{deph} \approx - \Gamma_{nm}^{(ch)} \, \rho_{nm}(\textbf{k},t) $, where $\Gamma_{nm}^{(ch)}$ is a constant rate that is proportional to the inverse of the core-hole lifetime.

The effects of electron-phonon couplings, which play a role at longer time scales, can also be approximated by using the electron–phonon Boltzmann scattering rate \cite{Haug2004}. In order to derive this scattering rate, the electron-phonon Hamiltonian needs to be considered, which enables to exchange energy and momentum between electrons and phonons. The additional coupling creates new terms in the equations of motion. Considering phonons as a thermal bath and performing also the Markov approximation, one arrives to the electron–phonon Boltzmann scattering rate, which contains the sum of terms in the bath that depend on the product of the density matrix in ${\bf k}$ and ${\bf k}+{\bf q}$, where ${\bf q}$ is the momentum transfer to a phonon.

\subsection{The initial state}

At zero temperature, all states below the Fermi energy level are assumed to be occupied, i.e. $\rho_{nn}({\bf k},t_0)=1$ if $\varepsilon^0_{n,{\bf k}} < \varepsilon_F$, being $\varepsilon_F$ the Fermi energy. The electron-electron interaction gives rise then to an energy that depends on the initial equilibrium state. This reference energy can naturally produce a constant energy shift in our initial bands. In order to correct for this effect, this energy is subtracted from the total Hamiltonian, or what is the same, the electron-electron interaction can be computed as $H_{e-e} (\textbf{k})\vert_{nm} = - \sum_{\mathbf{k}'} W_{n\textbf{k}'m\textbf{k},m\textbf{k}' n\textbf{k}} (\rho_{nm}({\bf k}',t)-\rho_{nm}({\bf k}',t_0)) $, in which the energy correlation arising from the equilibrium state is removed.

\subsection{The current}

By analogy to classical electromagnetism, the current generated by an electron is given by the product of its charge and velocity ${\bf J}=-\vert e \vert {\bf v}$. Hence, one can calculate the derivative of the mean value of the position for an electron in a periodic system, related to the current density, as
\begin{eqnarray} \label{Current}
\langle \hat{{\bf  J}} \rangle = - {\vert e \vert\over V} {d \over dt} \langle \hat{{\bf  r}} \rangle = -\frac{\vert e \vert}{i\hbar V} \langle [\hat{\bf r}, \hat{ H}_{sys}] \rangle
\end{eqnarray}
where $V$ is the volume of the system, $V=N \Omega_{uc}$, being $N$ the total number of unit cells and $\Omega_{uc}$ the volume of a unit cell. For a two-dimensional material, the volume would be substituted by area, i.e. $V=N A_{uc}$, where $A_{uc}$ corresponds to the area of a unit cell. All electrons involved in the dynamics, contribute to this formula. The position and the total Hamiltonian are single particle operators, and after some algebra, the current can be cast as
\begin{equation}
\langle \hat{{\bf  J}} \rangle = -\frac{|e|}{\hbar V}\sum_{{\bf k}\in \Omega_{bz}} \sum_{nm}  \Big( {\boldsymbol\nabla}_{\bf k} H_{nm}({\bf k}) -i [ {\boldsymbol \xi}, H]_{nm} ({\bf k})\Big) \rho_{mn}({\bf k})
\end{equation}
in which the total Hamiltonian matrix is $H({\bf k}) = H_0({\bf k}) + H_{e-e}({\bf k})$, without the light-matter interaction Hamiltonian, because it commutes with the position operator. Note also that the time dependence is both in the density matrix and in the excitonic interaction term. The previous formula has the same structure for any basis satisfying the Bloch theorem. Therefore the current can be calculated in a Bloch localized basis with no need to change to the eigenstate basis. The integration over the quasi-momentum should be restricted to the first Brillouin zone $\Omega_{bz}$.

The intraband current is defined in the eigenstate gauge as the part of the current that arise from the diagonal terms of the density matrix
\begin{equation}\label{Jintra}
\langle \hat{{\bf  J}}_{1} \rangle = -\frac{|e|}{\hbar V}\sum_{{\bf k}\in \Omega_{bz}} \sum_{n}  v_{nn}({\bf k}) \rho_{nn}({\bf k})
\end{equation}
where we define the velocity matrix as $v_{nm}({\bf k})= {\boldsymbol\nabla}_{\bf k} H_{nm}({\bf k}) -i [ {\boldsymbol \xi}, H]_{nm} ({\bf k})$. It is easy to find the the intraband current in the Bloch basis as
\begin{eqnarray}\label{JintraB}
\langle \hat{{\bf  J}}_{1} \rangle = -\frac{|e|}{\hbar V}\sum_{{\bf k}\in \Omega_{bz}}\sum_{n}\sum_{aa'bb'}  v_{nm}^{(B)}({\bf k}) U_{ma}({\bf k}) U^\dagger_{ab}({\bf k}) \nonumber\\ 
\times \; \rho_{bb'}^{(B)}({\bf k}) U_{b'a'}({\bf k}) U^\dagger_{a'n}({\bf k}) \nonumber \\
\end{eqnarray}

On the other hand, knowing the total and the intraband current, the interband current is just obtained by subtraction $\langle \hat{{\bf  J}}_{2} \rangle = \langle \hat{{\bf  J}} \rangle - \langle \hat{{\bf  J}}_{1} \rangle$.

\subsection{The optical absorption}

The induced dipole density of the system, or polarization of the system, is given by  
\begin{eqnarray}
\langle \hat{{\bf  P}} \rangle = - {\vert e \vert\over V} \langle \hat{{\bf  r}} \rangle 
\end{eqnarray}
where $- {\vert e \vert} \langle \hat{{\bf  r}} \rangle $ represents the dipole of all electrons and $V$ the volume of the quantum system. By analyzing the gain and loss of a quantum system under a laser pulse, one can relate the Fourier transform of the polarization with the absorption coefficient as \cite{Wu2016,Picon2019}
\begin{eqnarray}\label{Absorption}
\alpha(\omega) = \frac{\omega}{n_b c \epsilon_0}\frac{ {\rm Im} [{\bf P}(\omega)\cdot{\boldsymbol \varepsilon}^*(\omega)] } {\vert { \varepsilon}(\omega)\vert^2}
\end{eqnarray}
where $\epsilon_0$ is the dielectric permittivity of the vacuum, $n_b$ is the background refractive index, ${\boldsymbol \varepsilon}(\omega)$ is the Fourier transform of the external electric field ${\boldsymbol \varepsilon}(t)$, and  ${\bf P}(\omega) \equiv \int dt \; e^{-i\omega t} \langle \hat{{\bf  P}} \rangle (t)$. Note that defining the Fourier transform of the current as ${\bf J}(\omega) \equiv \int dt \; e^{-i\omega t} \langle \hat{{\bf  J}} \rangle (t) $, it is easy to check that $i\omega {\bf P}(\omega) = {\bf J}(\omega) $. The absorption coefficient has units of reciprocal length. For a two-dimensional material, because the volume is substitued by the area, then the absorption becomes unitless. In this last case, the material can be interpreted as a zero-thickness layer between two dielectric media, and one has to take into account the Fresnel coefficients for the reflection and the transmission accordingly \cite{Li2018}. The refractive index $n_b$ is considered to slowly change with $\omega$ within the spectral bandwidth of the laser pulse. { The absorption cross section is then $\sigma(\omega) = \alpha(\omega)/\Pi_c$ }, where $\Pi_c$ is the number of unit cells per volume. Note that the formula for the absorption given by Eq. (\ref{Absorption}) is nonperturbative and is also valid for a broad bandwidth pulse, which is suitable to describe attosecond pulses that may have a bandwidth of more than 10 eV.

In the optical linear response regime, the polarization is proportional to the applied electric field $P_i (\omega) = \chi_{ij}(\omega) \, \varepsilon_{j} (\omega)$, being $\chi_{ij}$ the susceptibility at first order. For a particular polarization direction $i$ ($=x,y,z$), the absorption coefficient is related to the imaginary part of the susceptibility by

\begin{eqnarray}
\alpha(\omega) = \frac{\omega}{n_b c \epsilon_0} {\rm Im} [\chi_{ii}(\omega)] 
\end{eqnarray}

If the coupling of the light with the material is weak, the absorption coefficient can be found at first-order time-dependent perturbation theory 

\begin{eqnarray} \label{Kubo}
\alpha(\omega)&=\frac{\pi |e|^2 \omega}{ c \epsilon_0 V }  \sum_{N}   |{\bf u} \cdot \braket{X_N|\hat{\bf r} |GS}|^2 \, \delta(\hbar \omega-[\epsilon_f-\epsilon_i]), \, \nonumber \\
\end{eqnarray}

where a set of final excited states given by single-particle excitations from valence (v) to conduction bands (c) is assumed, i.e.  $\ket{X_N}=\Sigma_{vc {\bf k}}A_{cv}^{(N)}({\bf k})\hat{c}_{c}^{\dagger}\hat{c}_{v}\ket{GS}$. The wavefunction $A_{cv}^{(N)}(\rm k)$ is typically found by solving the Bethe-Salpeter equation \cite{Rohlfing1998}. ${\bf u}$ is the polarization direction of the electric field. $\epsilon_f$ and $\epsilon_i$ are the energies of the final and initial states coupled by the laser light. In the absence of interactions, none of the electron-hole pairs are correlated and the absorption in the linear response is given by the Kubo-Greenwood formula

\begin{eqnarray}\label{Kubo-Greenwood}
\alpha(\omega)&=\frac{\pi |e|^2 \omega}{ c \epsilon_0 V }   \sum_{cv {\bf k}}|{\bf u} \cdot \braket{c {\bf k}|\hat{\bf r}_1 | v {\bf k}}|^2 \delta(\hbar \omega-[{\epsilon}_f({\bf k})-{\epsilon}_i({\bf k})]). \nonumber \\
\end{eqnarray}

In the linear regime the optical conductivity $\sigma_{ij}$ is related to the current $J_{i}(\omega) = \sigma_{ij}(\omega) \varepsilon_j (\omega)$, and consequently to the susceptibility by $i\omega \, \chi_{ij}(\omega) = \sigma_{ij}(\omega)$. More details are given in the appendix and in Ref. \cite{Esteve2022}

\section{\label{sec:numerics}Numerical implementation}

In this section we detail the numerical implementation for resolving the equations of motion for the density matrix of the system. These simulations require integration for hundreds of femtoseconds, implying an important computational effort, especially for three-dimensional materials that require a large grid. We design a code that resorts to both Message Passing Interface (MPI) and Open Multi-Processing (OMP) parallelization in order to speed up the simulations. In the following sections we describe the main features. 

\subsection{The grid in the reciprocal space}

The EOM for evolving the density matrix are defined in the reciprocal space. In order to reduce the computational effort, it is convenient to simulate the dynamics in the first Brillouin zone by also imposing periodic boundary conditions. In periodic systems, it is customary to work with Monkhorst-Pack grids \cite{Monkhorst1976} that simplifies the numerical implementation of boundary conditions for any given system with a particular spatial crystal symmetry. Any point in the reciprocal space can be written as ${\bf k} = k_x {\bf x} + k_y {\bf y} + k_z {\bf z}  = k_1 {\bf b}_1 + k_2 {\bf b}_2 + k_3 {\bf b}_3 $, where $(k_x,k_y,k_z)$ are cartesian coordinates in the direction of the canonical vectors $\{{\bf x},{\bf y},{\bf z}\}$, while $(k_1,k_2,k_3)$ are crystal coordinates in the direction of the reciprocal-lattice vectors $\{{\bf b}_1,{\bf b}_2,{\bf b}_3\}$. In crystal coordinates, it is easy to define the first Brillouin zone that spans from 0 to 1 in each coordinate. In each crystal direction, we take a discrete grid equally spaced. Any function that depends on the quasi-momentum $f({\bf k})$ is then represented by an array whose size is given by $N=N_1\times N_2 \times N_3$, which $N_i$ is the total number of points in the corresponding crystal direction. Because of periodic conditions, the last point $k_{N_i-1}$ in a particular direction must be effectively neighbor of the first one $k_{0}$ when we calculate the gradient or the excitonic interactions.

Note that once we know any function in crystal coordinates, we can represent the function in cartesian coordinates by using the  transformation
\begin{eqnarray} \label{crystal_coord}
\left( \begin{array}{ccc}
k_x  \\
k_y  \\
k_z 
\end{array} \right) = \left(
\begin{array}{ccc}
b_{1x} & b_{2x} & b_{3x}  \\
b_{1y} & b_{2y} & b_{3y}  \\
b_{1z} & b_{2z} & b_{3z}  
\end{array} \right) \left( \begin{array}{ccc}
k_1  \\
k_2  \\
k_3 
\end{array} \right)
\end{eqnarray}
where we define
\begin{equation*}
{\bf b}_i = b_{ix} {\bf x} +
b_{iy} {\bf y} +
b_{iz} {\bf z}
\end{equation*}
Note also that the discretization of the Brillouin zone can also be interpreted as the volume of the system, i.e. $V=N_1 N_2 N_3 \Omega_{uc}$.

Any integration with respect to the quasi-momentum in cartesian coordinates is translated to crystal coordinates by using the Jacobian determinant $\sum_{k_xk_yk_z} \rightarrow \det(M) \sum_{k_1k_2k_3}$, where { $M$ is the matrix of the transformation (\ref{crystal_coord}).}

\subsection{The multidimensional arrays}
The density matrix and the different Hamiltonians depend on three parameters: the quasi-momentum ${\bf k}$ and the two band quantum numbers $m$ and $n$. We represent them by multidimensional arrays with the structure $f[k][m][n]$. Here, $k$ is an integer that is mapped to a reciprocal point ${\bf k}$ of the grid. In many loops, in order to calculate observables or computing the EOM, we run over these three parameters. As $k$ is typically the parameter with more points, the external loop is always with $k$, while the internal loop is with $n$. This enables us to use multi threading in the $k$ loop. We create special multidimensional arrays for $f[k][m][n]$ whose data is aligned. This enables us to use autovectorization in the internal $n$ loop, which does not compromise the multi threading in the external loop.

The Berry connections $\boldsymbol{\xi}({\bf k})$ and the non-interacting Hamiltonian $H_0({\bf k})$ can be calculated with any density functional theory or hartree-fock code that allows one to express them in a localized Bloch basis, as given in Eq. (\ref{Bloch_basis}), such as CRYSTAL or SIESTA \cite{CRYSTAL,SIESTA}. Because in these basis they are smooth in ${\bf k}$, we use a coarse grid to calculate them and then we interpolate them in order to obtain a finer grid for resolving the time evolution, see Ref. \cite{Esteve2022} for more details of how to calculate the Berry connections. Similarly, we could use tight-binding models to provide the Berry connection and the non-interacting Hamiltonian via an analytical expression. Also, it is possible to calculate these elements by using localized Wannier orbitals \cite{Marzari2012,Rui2019}, which can be generated from codes such as Wannier90 \cite{Wannier90}. In the next section we show some examples in which we successfully implement different schemes based on Bloch, tight-binding, and Wannier basis.

\subsection{The gradient in the EOM}

The discretization of the reciprocal space in a Monkhorst-Pack grid requires a proper definition of the numerical gradient: the functions are not known at points spanned over the cartesian axes, but over non-orthogonal directions defined by the reciprocal lattice vectors.

A numerical implementation of the gradient in a Monkhorst-Pack grid has been extensively used in electronic structure calculations in equilibrium systems \cite{Marzari2012}. The implemented gradient has a linear order precision \cite{Mostofi2008}. However, computing the the equations of motion and resolving the non-equilibrium dynamics requires a higher precision in order to reach convergence. 

\begin{figure} 
  \includegraphics[width=0.49\textwidth]{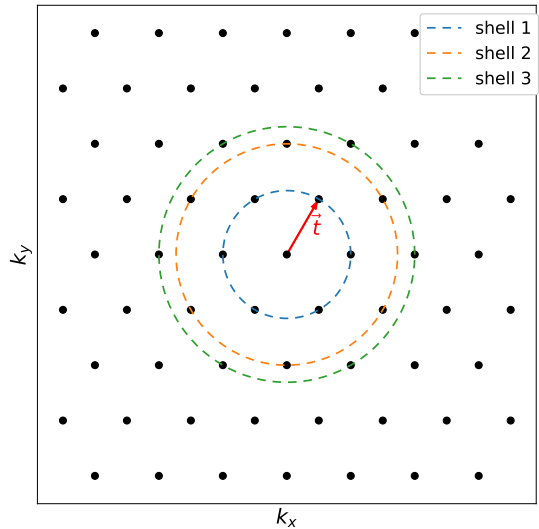}
  \caption{Monkhorst-Pack grid of the reciprocal space for a two-dimensional material with hexagonal symmetry. The calculated gradient at the point located at the origin depends on the neighbor points. Those are divided in shells, which are defined by the distance to the origin.  
  }\label{fig:gradient}
\end{figure}

We extend the method used in \cite{Mostofi2008} up to cubic order precision. The main difference lies on the constraints of the neighbors points to be satisfied. We define the vectors $\textbf{t}$ that connect each $\textbf{k}$ point of the grid with its closest neighbors. The $\textbf{t}$ vectors are divided in shells, ordered by length, see Fig. \ref{fig:gradient} for a two-dimensional hexagonal lattice. The gradient of a smooth function $f({\bf k})$ is calculated as:
\begin{eqnarray} \label{gradient}
\nabla_{\mathbf{k}} f(\mathbf{k})=\sum_{\mathbf{t} \in s} \omega_{s} \mathbf{t}[f(\mathbf{k}+\mathbf{t})-f(\mathbf{k})]
\end{eqnarray}
where $\omega_s$ is the weight for a particular shell $s$. If the $s$-th shell contains $M_s$ vectors and the total number of shells is $N_s$, the constraints to compute the gradient are
 \begin{equation}\label{constraint1}
    \sum_{s=1}^{N_s} \omega_s \sum_{i=1}^{M_s} t_\alpha^{i} t_\beta^i =\delta_{\alpha\beta}
\end{equation}
\begin{equation}\label{constraint2}
    \sum_{s=1}^{N_s} \omega_s \sum_{i=1}^{M_s} t_\alpha^{i} t_\beta^i t_\gamma^i t_\eta^i =0
\end{equation}
Both equations need to be satisfied simultaneously. The Greek labels refer to the ${\bf t}$ vector components in the $x$, $y$, and $z$ directions. Note that Eq. (\ref{constraint1}) and Eq. (\ref{constraint2}) give rise to 6 and 15 different equations, so we have in total 21 independent equations. We can rewrite the constraints in a matrix form 
\begin{equation}\label{system}
A {\boldsymbol \omega }= {\bf q}
\end{equation}
where A is a matrix of dimensions $21 \times N_s$, whose first six rows are:
\begin{equation*}
    \sum_{i=1}^{M_s} t_\alpha^{i} t_\beta^i, 
\end{equation*}
and the remaining 15 rows are:
\begin{equation*}
     \sum_{i=1}^{M_s} t_\alpha^{i} t_\beta^i t_\gamma^i t_\eta^i 
\end{equation*}
while $\boldsymbol{\omega}$ and ${\bf q}$ are both vectors of length $N_s$ and 21 respectively. The $\boldsymbol{\omega}$ vector contains the weights of the shells, which are the ones to be calculated in order to be able to compute the gradient with the formula given by Eq. (\ref{gradient}). The ${\bf q}$ vector contains the information on the right hand of the constraints (\ref{constraint1}) and (\ref{constraint2}).

We construct the A matrix, for each shell of increasing length, and we try to invert the system Eq. (\ref{system}) by using the Moore-Penrose pseudoinverse of A. The algorithm runs until the weights found by using the pseudoinverse of A, i.e. $\boldsymbol{\omega} = A^{-1} {\bf q}$, satisfy Eq. (\ref{system}).

\subsection{Parallelization of the $\textbf{k}$ grid}

\begin{figure} 
\includegraphics[width=0.49\textwidth]{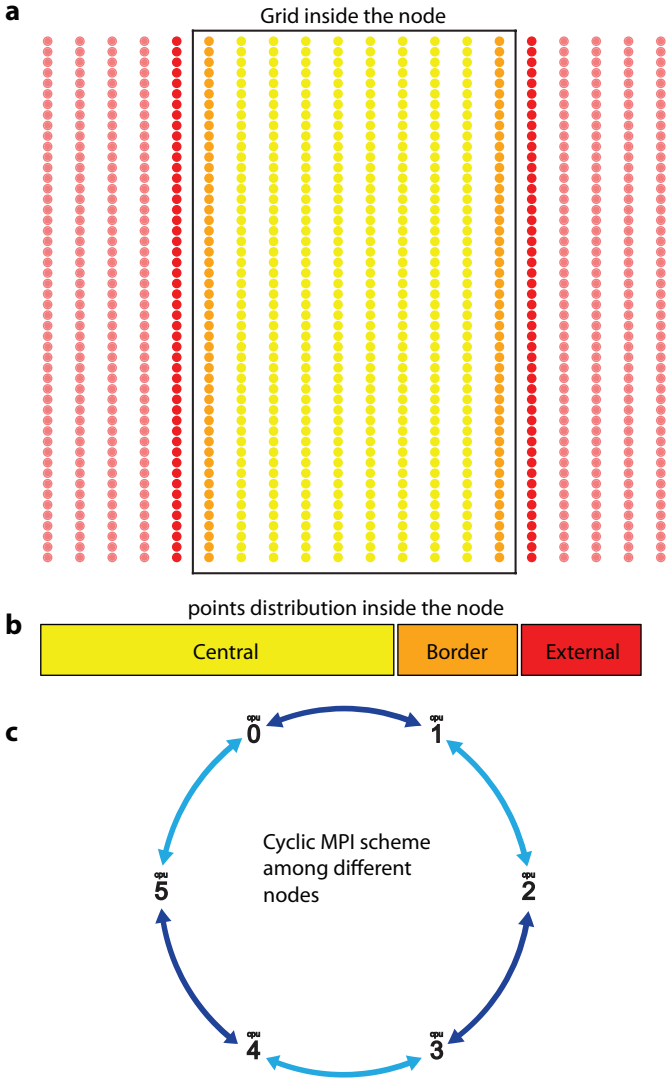}
\caption{MPI communication scheme. a) A two-dimensional grid in reciprocal space, in crystal coordinates, that is split in different nodes along the direction of one reciprocal-lattice vector. b) Inside a node, the ${\bf k}$ points are classified in central (yellow), border (orange), and external (red) points. The yellow part is propagated internally, the border part is passed to other nodes, and the external part is read from other nodes. c) MPI communication scheme. Each node communicate with two different ones, in a cyclic way as defined in the picture.
}\label{fig:mpi}
\end{figure}

The reciprocal space is divided in different nodes by using MPI libraries, see figure \ref{fig:mpi}. Each node load their corresponding initial parameters, such as non-interacting Hamiltonian and Berry connections, depending on their position in ${\bf k}$. Hence, our multidimensional arrays $f[k][m][n]$ can be distributed in different nodes and enables us to perform simulations that require a high demand of memory. In each node, multi threading and autovectorization are implemented as previously described.

The gradient in ${\bf k}$ of the density matrix is part of the light-matter interaction term and it is important for evolving the density matrix, see Eq. (\ref{EOM}). The gradient of the non-interacting Hamiltonian is important to calculate the current, see Eq. (\ref{Current}), although this term is not time-dependent. The implementation of the gradient requires knowledge of the neighbor ${\bf k}$ points, see Eq. (\ref{gradient}). This requires to set an MPI scheme to communicate this information among nodes, since each node is in charge of propagating only a part of the density matrix, there will be some ${\bf k}$ points whose neighborhood is not inside the same node. In particular, we identify the ${\bf k}$ points in three different categories, see figure \ref{fig:mpi}:
\begin{itemize}
\item Central points: in those points the density matrix is evolved inside the node and we do not need to transfer this information to other nodes.
\item Border points: in those points the density matrix is evolved inside the node and we need to transfer this information to other nodes.
\item External points: in those points the density matrix is not evolved inside the node. The information of the density matrix at those points needs to be received for the propagation of other points.
\end{itemize}

To minimize the MPI communication, a good choice for splitting the reciprocal space is to cut in the direction defined by one of the reciprocal-lattice vectors: this way the external points will only lay on lines at the edges, see figure \ref{fig:mpi}a. This defines the different points of the grid inside the node as central, border, and external points. The communication is then cyclic for the nodes because the reciprocal space splitting is defined on a periodic direction, see figure \ref{fig:mpi}c.

At each step of the propagation, the density matrix located at the border and the external points are passed from one node to the ones within the neighborhood. This is the source of some overhead time. In the next section we show some particular examples for two- and three-dimensional materials. The MPI scheme scales well for low number of nodes, approaching the $\frac{1}{\# nodes}$ optimal law. The time propagation is implemented by a Runge-Kutta method of fourth order. Other time propagators are also implemented using Euler methods. 

\subsection{The mean-field electron-electron interaction}

In the mean-field approximation, the electron-electron interaction Hamiltonian becomes an additional time-dependent term to the non-interacting Hamiltonian, see Eq. (\ref{HeeMF}). This involves a sum over the ${\bf k}$-space of the density matrix times the Coulomb interaction $W_{n\textbf{k}'m\textbf{k},m\textbf{k}' n\textbf{k}}$, whose definition is a two-particle integral (\ref{CoulombW}). For Bloch wave functions $\psi_{n,\bf k}(\textbf{r}) = e^{i\textbf{k}\cdot \textbf{r}} u_{n,\textbf{k}}(\textbf{r})$, the Coulomb interaction can also be expressed as \cite{Ridolfi2020}
\begin{eqnarray}
W_{n\textbf{k}'m\textbf{k},l\textbf{k}' j\textbf{k}} =   \sum_{\textbf{G}}\left[I_{j\textbf{k},n\textbf{k}'}^\textbf{G}\right]^*I_{m\textbf{k}, l\textbf{k}'}^{\textbf{G}} \; V_{\textbf{k}'-\textbf{k}+\textbf{G}} \nonumber\\
\end{eqnarray}
where ${\bf G}$ is a sum over reciprocal-lattice vectors, $V_{\bf k}$ is the Fourier transform of the Coulomb energy $V({\bf r})$, and 
\begin{eqnarray*}
I_{m\textbf{k}, m'\textbf{k}'}^\textbf{G} &\equiv&\int d^3 r \; u^*_{m\textbf{k}}(\textbf{r}) u_{m'\textbf{k}'}(\textbf{r})e^{-i\textbf{G}\cdot \textbf{r}}\\
&=&\int d^3 r \; \psi^*_{m\textbf{k}}(\textbf{r}) \psi_{m'\textbf{k}'}(\textbf{r})e^{i(\textbf{k}-\textbf{k}'-\textbf{G})\cdot \textbf{r}}
\end{eqnarray*}

We can further expand the integral $I_{m\textbf{k}, m'\textbf{k}'}^\textbf{G}$ by using a localized Bloch basis (\ref{Bloch_basis}) via the unitary transformation of Eq. (\ref{unitary}) and reduce the integral to

\begin{eqnarray*}
I_{m\textbf{k}, m'\textbf{k}'}^\textbf{G} &\approx&\sum_{\alpha} U_{\alpha m}^* (\textbf{k}) U_{\alpha m'} (\textbf{k}')  e^{i(\textbf{k}-\textbf{k}'-\textbf{G})\cdot \textbf{t}_\alpha} 
\end{eqnarray*}
where we have used the approximation of very localized orbitals. Here ${\bf t}_\alpha$ is the position of the atom where the orbital is localized. Within this approximation, the electron-electron Hamiltonian in a Bloch basis $H_{e-e}^{(B)}({\bf k})$ is simplified as
\begin{eqnarray} \label{HeeBloch}
 H_{e-e}^{(B)}  (\textbf{k})\vert_{nm} && = \nonumber \\
&-& \!\! \sum_{\mathbf{k}',{\bf G}} e^{i(\textbf{k}'-\textbf{k}-\textbf{G})\cdot (\textbf{t}_{m}-\textbf{t}_n)} V_{\textbf{k}-\textbf{k}'+\textbf{G}} \; \rho_{nm}^{(B)}({\bf k}',t) = \nonumber \\
&-& \!\! \sum_{\mathbf{k}'} \tilde{V}_{\textbf{k}-\textbf{k}'}^{nm} \; \rho_{nm}^{(B)}({\bf k}',t) 
\end{eqnarray}

From the numerical viewpoint, there is an advantage to work in a Bloch basis expanded in localized orbitals in order to avoid large grids for the reciprocal space. The expression given by Eq. (\ref{HeeBloch}) is extremely convenient to calculate the electron-electron interactions. Note that the density matrix evolves in time, therefore this Hamiltonian also needs to be computed at each time step. Note that the effective potential $\tilde{V}_{\textbf{k}-\textbf{k}'}^{nm} = \sum_{\bf G} e^{i(\textbf{k}'-\textbf{k}-\textbf{G})\cdot (\textbf{t}_{m}-\textbf{t}_n)} V_{\textbf{k}-\textbf{k}'+\textbf{G}} $ is periodic in the reciprocal space. For the rest of the manuscript we use $\tilde{V}_{\textbf{q}}^{nm} \equiv \tilde{V}_{\textbf{q}}$ to simplify the notation, but always keeping in mind that there is a phase factor that depends on the band indexes.

The Fourier transform of the Coulomb energy between electrons $V(r) = e^2/4\pi\epsilon_0 r$ is given by \cite{Haug2004} 
\begin{eqnarray} \label{CouPot}
V_{\bf q} = V_{\bf q}^{(3D)} = \frac{e^2}{\epsilon_0 V} \frac{1}{ |{\bf q}|^2} ,\, V_{\bf q} = V_{\bf q}^{(2D)} = \frac{e^2}{2 \epsilon_0 A} \frac{1}{ |{\bf q}|} ,\;\;
\end{eqnarray}
for a 3D and 2D material. However, in the case of two-dimensional materials, due to dielectric effects of the environment and screening, it is more convenient to use the so-called Rytova-Keldysh potential \cite{Rytova1967,Keldysh1979}
\begin{eqnarray}
V_{RK}(r)=\frac{1}{4\pi\epsilon_0}\frac{\pi e^{2}}{(\epsilon_1+\epsilon_2) r_{0}}\left[H_{0}\left(\frac{r}{r_{0}}\right)-Y_{0}\left(\frac{r}{r_{0}}\right)\right]
\end{eqnarray}
which is a 2D electrostatic potential derived for a thin layer embedded between two dielectrics. $H_0$ and $Y_0$ are the Struve function and the Bessel function of the second kind. The screening length is $r_0 = d  \,\epsilon_m/(\epsilon_1 + \epsilon_2)$, where $d$ is the thickness of the material, and $\epsilon_m$, $\epsilon_1$ and $\epsilon_2$ are the dielectric constants of the material, the top dielectric medium, and the bottom dielectric medium, respectively. The Fourier transform of the Rytova-Keldysh potential is given by
\begin{eqnarray} \label{RKpot}
V_{\bf q} = V_{\bf q}^{(RK)} = \frac{e^2}{\epsilon_{0}(\epsilon_1+\epsilon_2)A} \frac{1}{ |{\bf q}|\left(r_{0} |{\bf q}|+1 \right)}
\end{eqnarray}
In the case of a free-standing monolayer, then $\epsilon_1=\epsilon_2=1$. 

The dynamical mean-field interaction (\ref{HeeMF}) implies a high computational cost. First of all, for a grid with $N$ points in the BZ it requires to perform $N^2$ operations each time step. Second, the computation of this sum around ${\bf k} \approx {\bf k}'$ requires more accurate evaluation due to the sharpness of the Coulomb interaction. In the following we discuss our numerical implementation in order to avoid the above mentioned difficulties and exploit at the same time parallelization resources.

We can always write the effective potential of Eq. (\ref{HeeBloch}) as $\tilde{V}_{\bf q} = \tilde{V}_{\bf q}^{(s)} + \Delta \tilde{V}_{\bf q}$, where $\tilde{V}_{\bf q}^{(s)}$ is a smooth potential around ${\bf q}=0$ and $\Delta \tilde{V}_{\bf q}$ contains the singularity at ${\bf q}=0$. In order to construct $\tilde{V}_{\bf q}^{(s)}$, we change $\vert {\bf q} \vert \rightarrow \sqrt{q^2 + q^2_{TF}}$ in $V_{\bf q}$, where $q_{TF}$ is a Thomas-Fermi screening parameter. This parameter makes the potential smooth around the origin, and in the limit of $q_{TF}\rightarrow 0 $ we recover then the singular potential. We therefore define $\Delta \tilde{V}_{\bf q} = \tilde{V}_{\bf q}- \tilde{V}_{\bf q}^{(s)} $, which is close to zero at long range and presents a singularity at close range. We can always choose a small $q_{TF}$ such that $\tilde{V}_{\bf q}^{(s)}$ reproduces well the initial potential in all ${\bf k}$-space besides in a small area around $q \approx 0$. In the following we focus on two-dimensional systems, but the same procedure can be extended to three-dimensional ones. We expand the smooth function $\tilde{V}_{\bf q}^{(s)}$ in Fourier series 
\begin{eqnarray} \label{CouPot_Fourier}
\tilde{V}_{\bf q}^{(s)} = \sum_{u,v=0}^{N_{\rm cut}} [A_{uv} \cos(2\pi uq_x)\cos(2\pi vq_y) + \nonumber \\
B_{uv} \cos(2\pi uq_x)\sin(2\pi vq_y) + \nonumber \\
C_{uv} \sin(2\pi uq_x)\cos(2\pi vq_y) + \nonumber \\
D_{uv} \sin(2\pi uq_x)\sin(2\pi vq_y) ],
\end{eqnarray}
where $A_{uv}, B_{uv}, C_{uv}, D_{uv}$ are the Fourier coefficients, and $N_{\rm cut}$ is the highest order harmonic that we take into account. The smooth potential (\ref{CouPot_Fourier}) is periodic and it is relatively small at the boundaries of the first Brillouin zone. The dynamical mean-field interaction for the smooth potential is obtained by including Eq. (\ref{CouPot_Fourier}) in (\ref{HeeBloch})
\begin{eqnarray} \label{HeeMFs}
H_{e-e} ({\bf k})\vert_{nm}^{(s)} &= -\sum_{u,v=0}^{N_{cut}} [X^{cc}_{uv} \cos(2\pi uk_x)\cos(2\pi vk_y) +  \nonumber \\
&X^{cs}_{uv} \cos(2\pi uk_x)\sin(2\pi vk_y) + \nonumber \\
&X^{sc}_{uv} \sin(2\pi uk_x)\cos(2\pi vk_y) + \nonumber \\
&X^{ss}_{uv} \sin(2\pi uk_x)\sin(2\pi vk_y)]
\end{eqnarray}
$X_{uv}$ are coefficients that do not depend on ${\bf k}$ and are expressed as
\begin{eqnarray}
  X^{cc}_{uv} = A_{uv} f^{cc}_{uv} + B_{uv} f^{cs}_{uv} + C_{uv} f^{sc}_{uv} + D_{uv} f^{ss}_{uv}  &, \nonumber \\
  X^{cs}_{uv} = A_{uv} f^{cs}_{uv} - B_{uv} f^{cc}_{uv} + C_{uv}f^{ss}_{uv} - D_{uv} f^{sc}_{uv}  &, \nonumber \\
  X^{sc}_{uv} = A_{uv} f^{sc}_{uv} + B_{uv} f^{ss}_{uv} - C_{uv} f^{cc}_{uv} - D_{uv} f^{cs}_{uv}  &, \nonumber \\
  X^{ss}_{uv} = A_{uv} f^{ss}_{uv} - B_{uv} f^{sc}_{uv} - C_{uv} f^{cs}_{uv} + D_{uv} f^{cc}_{uv} &,
\end{eqnarray}
where
\begin{eqnarray}
  f^{cc}_{uv} = \sum_{\mathbf{k}} \rho^{(B)}_{nm}(\mathbf{k})\cos(2\pi uk_x)\cos(2\pi vk_y),    & \nonumber \\
  f^{sc}_{uv} = \sum_{\mathbf{k}} \rho^{(B)}_{nm}(\mathbf{k})\sin(2\pi uk_x)\cos(2\pi vk_y),    & \nonumber \\
  f^{cs}_{uv} = \sum_{\mathbf{k}} \rho^{(B)}_{nm}(\mathbf{k})\cos(2\pi uk_x)\sin(2\pi vk_y),    & \nonumber \\
  f^{ss}_{uv} = \sum_{\mathbf{k}} \rho^{(B)}_{nm}(\mathbf{k})\sin(2\pi uk_x)\sin(2\pi vk_y).
  \label{sums_rho_k_vector}
\end{eqnarray}
The coefficients $A_{uv}$, $B_{uv}$, $C_{uv}$ and $D_{uv}$ can be computed at the beginning. However, the functions $f_{uv}$ must be computed at each time step, as they depend on the density matrix. The sums over ${\bf k}$ in Eqs. (\ref{sums_rho_k_vector}) are performed in the first Brillouin zone. Note now that the number of operations to compute the mean-field interaction is $N\times 4 N_{\rm cut}^2$. Hence, if the number of coefficients in the Fourier series is small in comparison with the number of grid points in the reciprocal space, which is the typical case, then this is already an advantage. Because the {\bf k} grid is split in our parallelization scheme, see previous section, each node requires to have information of the time-dependent $X_{uv}$ coefficients to compute the mean-field energy (\ref{HeeMFs}). For that, each node needs to compute their corresponding sum in Eq. (\ref{sums_rho_k_vector}) and communicate this number to the rest of the nodes in order to calculate the $X_{uv}$ coefficients. Due to the small size message communication, this approach is indeed efficient, even though the $X_{uv}$ coefficients require to be computed at each time step.

\begin{figure} 
  \includegraphics[width=0.5\textwidth]{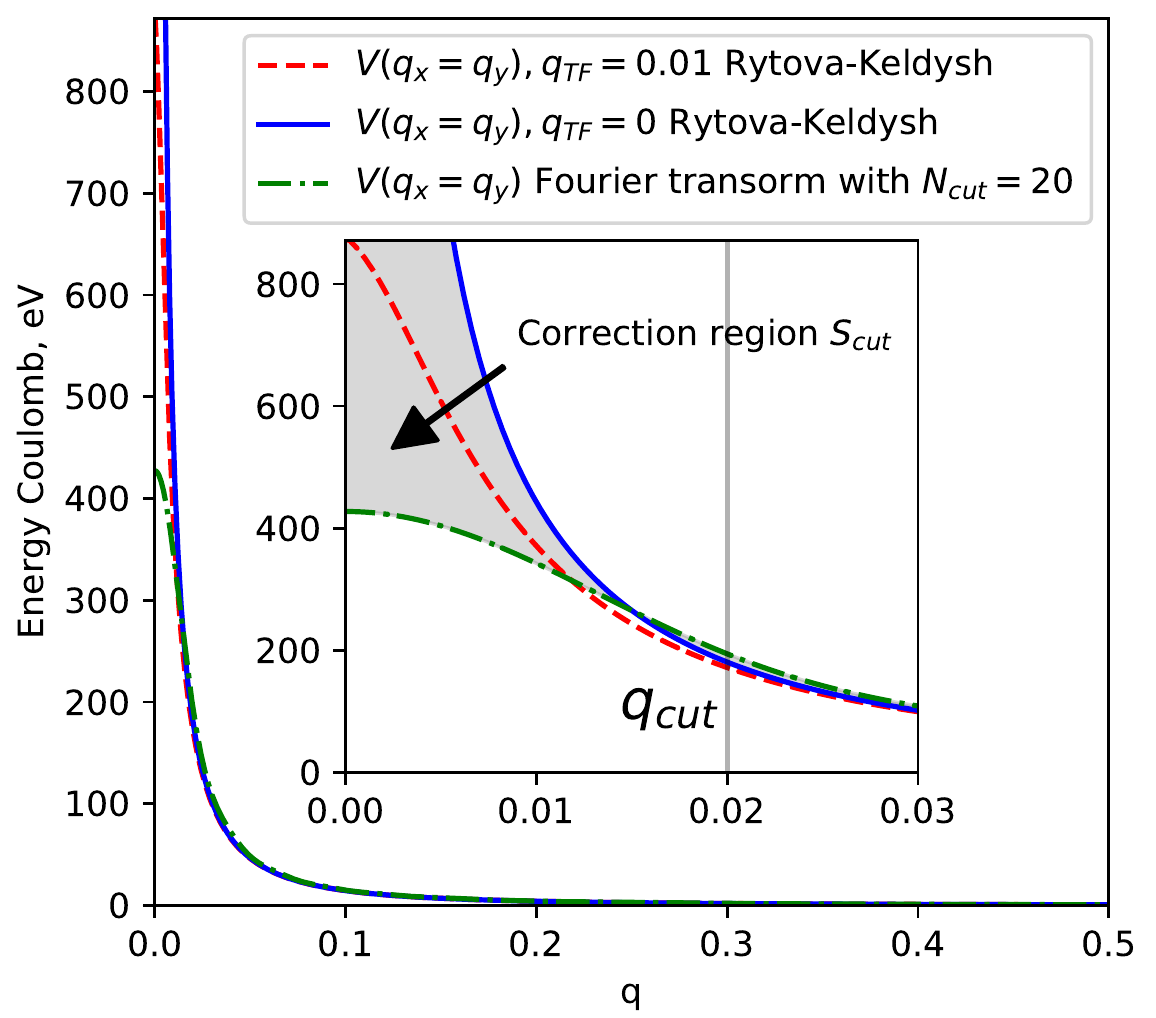}
  \caption{Comparison of exact Rytova-Keldysh potential $\tilde{V}_{\bf q}$ (blue line) and the same potential accounting for a Thomas-Fermi screening $q_{TF}$, $\tilde{V}_{\bf q}^{(s)}$ (red line). The expansion of the $\tilde{V}_{\bf q}^{(s)}$ potential in Fourier series (green line), with $N_{\rm cut}=20$, is practically the same, the main difference is around the origin. The vertical black line shows the position of $q_{\mathrm{cut}}$. The inset is a zoom around the origin q=0. The Rytova-Keldysh parameters are $A=\sqrt{3}a^2/2$, $a=2.5$ \AA, $r_0=10$ \AA, and $\epsilon_1=\epsilon_2=1$.
  }\label{fig:RK}
\end{figure}

The exact interaction term can be computed as $H^{(B)}_{e-e} ({\bf k})\vert_{nm} = H_{e-e} ({\bf k})\vert_{nm}^{(s)} + X^{corr}_{nm}(\mathbf{k},t)$, where $X^{corr}_{nm}(\mathbf{k},t)$ is a correction at close range due to the Thomas-Fermi screening parameter
\begin{equation}
X^{corr}_{nm}(\mathbf{k},t) 
= -\sum_{\mathbf{q}, \ q<q_{\mathrm{cut}}}   \left[ \tilde{V}_{\bf q}-\tilde{V}_{\bf q}^{(s)} \right]   \rho^{(B)}_{nm}(\mathbf{k+q},t)
\end{equation}
Because the correction is mainly localized around ${\bf q}=0$, then the sum over ${\bf q}$ is only meaningful around the origin, see for example Fig. \ref{fig:RK} for a Rytova-Keldysh potential in a monolayer boron nitride. Hence, we expand the density matrix around the origin $\rho_{nm}(\mathbf{k+q},t) = \rho_{nm}(\mathbf{k},t) + {\bf q}\cdot \partial_{\bf k}\rho_{nm}(\mathbf{k},t) + O(q_{cut}^2)$. Due to the symmetry of the potential, the linear terms vanish and we can approximate the correction as
\begin{equation}
X^{corr}_{nm}(\mathbf{k},t) \approx -\rho^{(B)}_{nm} (\mathbf{k},t) S_{\mathrm{cut}} + O(q_{cut}^2),
\end{equation}
where 
\begin{equation*}
   S_{\mathrm{cut}} = \sum_{\mathbf{q}, \ q<q_{cut}}   \left[ \tilde{V}_{\bf q}-\tilde{V}_{\bf q}^{(s)}
\right]
\end{equation*}
Note that the correction at a particular point is given by the density matrix at that same point times the factor $S_{cut}$, which can be computed at the beginning and does not depend on time.

\section{\label{sec:examples}Relevant examples}

In this section we show three relevant examples for our real-time electron dynamics code. First, the calculation of light-induced current and optical/UV absorption in a monolayer boron nitride (hBN). These calculations are performed for both a TB model and a Kohn-Sham (KS) Hamiltonian obtained from CRYSTAL code, which uses Gaussian-type local orbitals as a basis. Second, we show the extension of these calculations when excitonic interactions are considered. Third, we show the calculation of an ATAS spectrum for realistic parameters in a pump-probe experiment in graphite.  

\begin{figure} 
\includegraphics[width=0.49\textwidth]{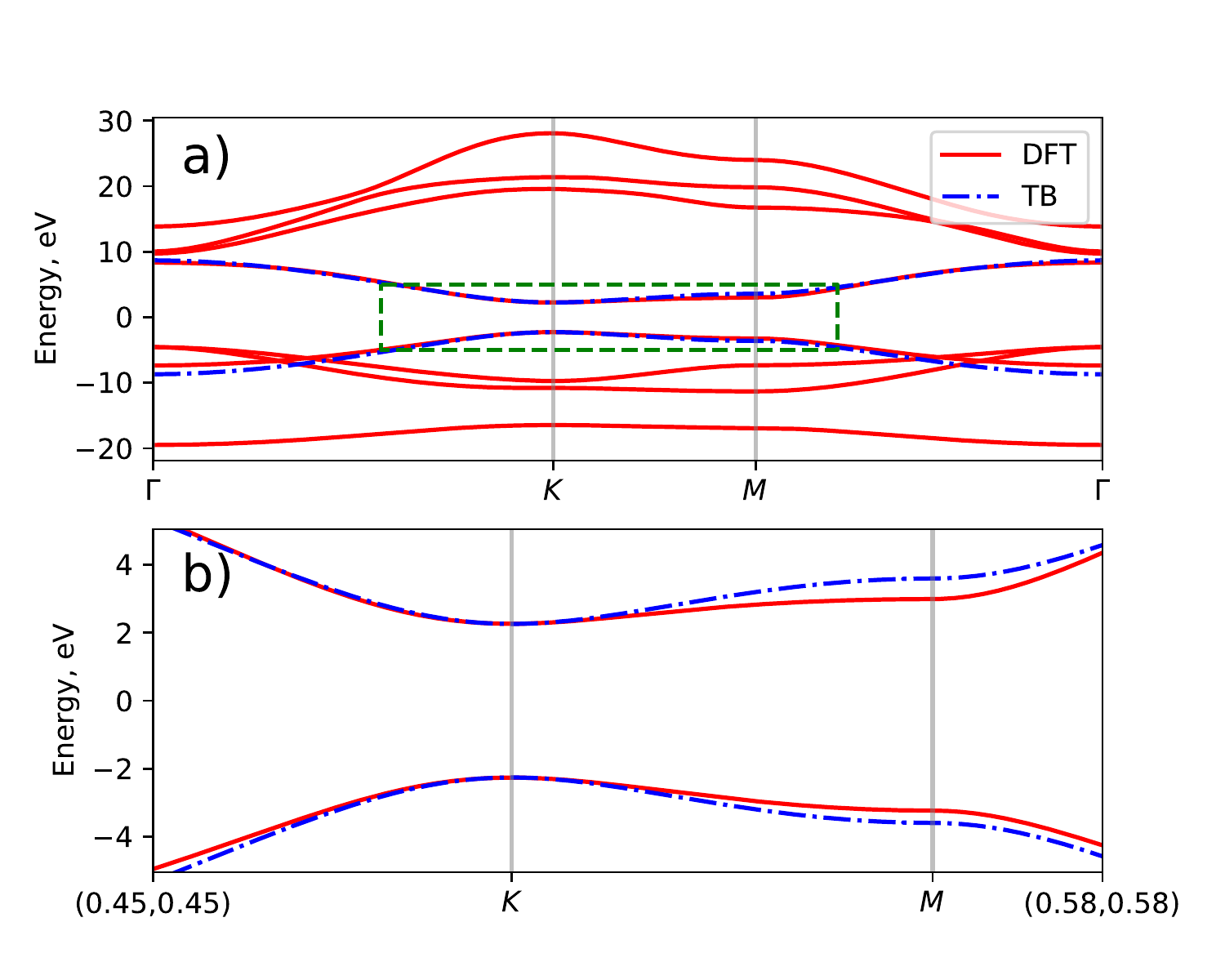}
  \caption { a) Band structure of hBN along the path $\Gamma - K - M - \Gamma$. The TB parameters have been found by fitting the energy dispersion around the $K$ point. b) Zoom in of the band energies around the $K$ point marked by the green rectangle.
  }\label{fig:bands_hBN}
\end{figure}

\begin{figure*}
\includegraphics[width=0.9\textwidth]{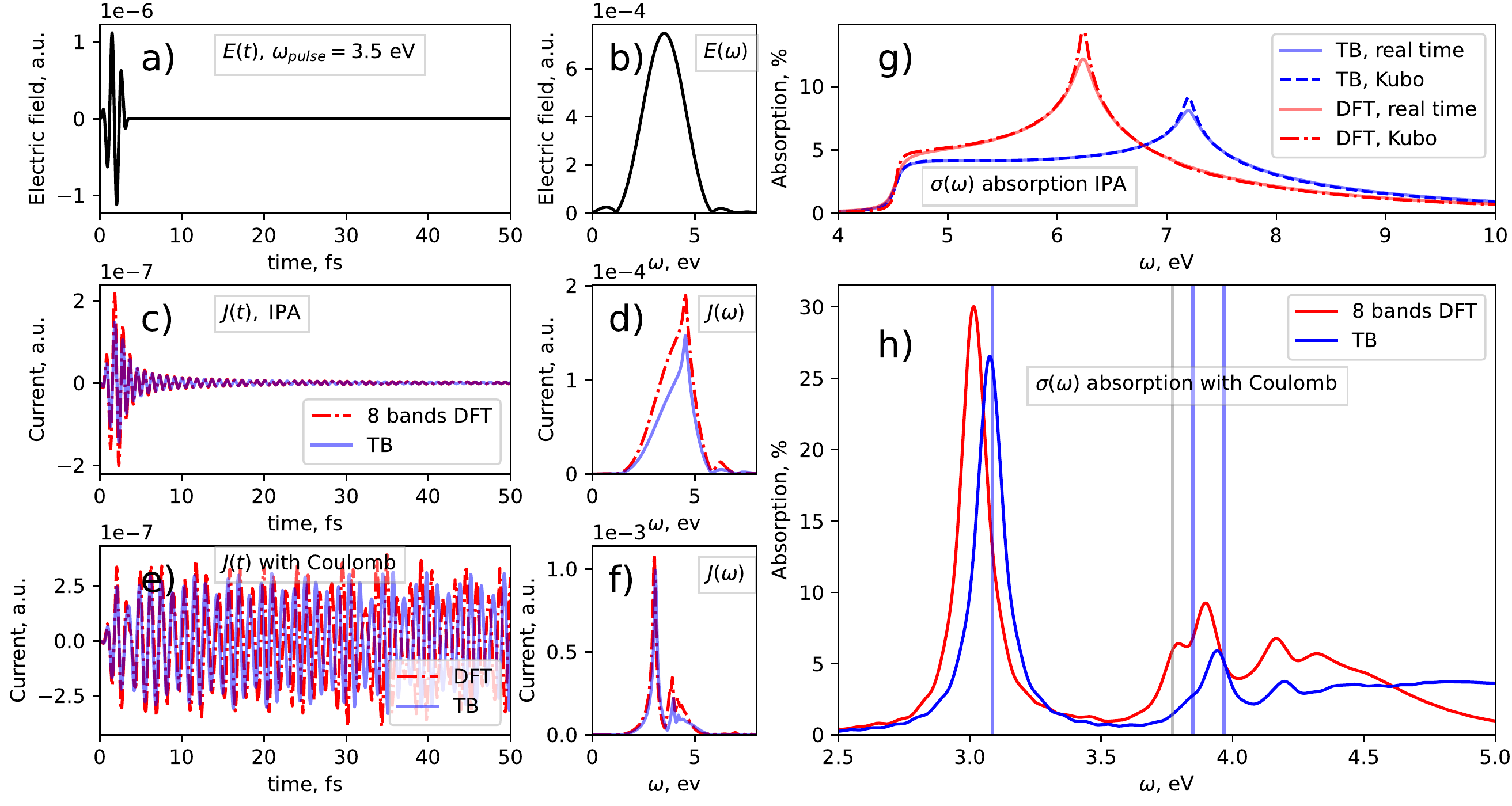}
\caption{ Current and absorption calculations in hBN. Figures a) and b) is the laser pulse in time and frequency domain. Figures c) and e) is the light-induced current without (independent particle approximation, IPA) and with excitons. Red dash-dotted lines represent the results for a 8 bands model obtained by DFT calculations, while solid blue lines are for a 2 band tight-binding model. Figures d) and f) are the corresponding Fourier transform. Figures g) and h) are the calculated absorption spectra without and with excitons. The real-time calculations are computed with Eq. (\ref{Absorption}) and compared with the first-order Kubo formula. The reciprocal space grid is $N=300\times300$ points, the number of terms in the Fourier series is $N_{\mathrm{cut}}=20$, and the time step is 1.2 as.
  }\label{fig:Abs}
\end{figure*}

\subsection{Current and optical/UV absorption in hBN}

Monolayer boron nitride is a material with hexagonal symmetry and a large optical bandgap of more than 4 eV. We first perform a DFT calculation within the LDA approximation using the CRYSTAL code \cite{CRYSTAL}. The unit cell has one atom of boron and nitrogen. A small number of Gaussian basis, one $s$- and three $p$-orbitals per atom, is enough to properly describe the energy structure around the bandgap, see Fig. \ref{fig:bands_hBN}. The bandgap energy is 4.5 eV, lower, as expected, than the experimental value \cite{Henriques2020}, but this is not essential to demonstrate the validity of our calculations. After calculating the Kohn-Sham electronic structure, we find the tight-binding model that fits the band structure around the $K$ points, see Fig. \ref{fig:bands_hBN}b.

In this first example, we neglect electron-electron interactions described by $H_{e-e} (\textbf{k})$ in eq. \ref{EOM}. First, we calculate the light induced current of the material by an ultrashort laser pulse. The time profile of the pulse is modeled by a $\sin^2$ envelope and a carrier wave of 3.5-eV frequency as shown in figure \ref{fig:Abs}a. The pulse intensity is weak, 10$^5$ W/cm$^2$, in order to be at the first-order perturbation regime. The pulse is polarized along the $\Gamma$-$M$ direction. The current is represented in figure \ref{fig:Abs}c. Note that even after the pulse, the current keeps oscillating and damps out after a few femtoseconds. The current for the TB model and the KS Hamiltonian show a very similar behavior. Second, we calculate the absorption of the laser pulse. The absorption is calculated within the bandwidth of the pulse, which it is around 4 eV centered at 3.5 eV, see \ref{fig:Abs}b. In order to calculate the absorption for other energies, such as for the highest energy point around 10 eV (the $\Gamma$ point), we calculate the current for different pulse frequencies. Because the bandwidth is broad, few pulses are enough to cover the whole spectrum. The resulting absorption for the TB and KS Hamiltonians are shown in figure \ref{fig:Abs}g. The absorption increases after the bandgap and the lineshape is similar for both models. However, we find a clear difference at higher energies. The peak of the absorption is located at the $M$ points, which are van Hove singularities. Those are located at different energies in the two models, see Fig. \ref{fig:bands_hBN}b.

In order to demonstrate the validity of the absorption calculation, we compare the results by using the Kubo-Greenwood formula (based on first-order perturbation theory) for a monochromatic laser given by equation (\ref{Kubo-Greenwood}). The Kubo formula is the standard method to obtain single-photon absorption spectra. The comparison is shown in Fig. \ref{fig:Abs}g. We observe that both procedures lead to similar results, apart from some smoothing in the absorption line of the real-time calculations due to the effects of a broad bandwidth pulse. This demonstrates the possibility to calculate optical/UV absorption spectra at the weak-intensity regime via real-time simulations. Furthermore, we could explore the non-linear regime beyond first-order perturbation theory by increasing the intensity or we could introduce a second pulse in order to model an ultrafast experiment, as we show in the last example for graphite.

The calculations were performed in Intel Xeon CPUs E5-2630 type with 2.40 GHz. The size of the reciprocal space grid is $N=300\times300$ points. The time step is controlled dynamically and varies in the range of 1.2 - 0.3 as during the simulation up to 80 femtoseconds. The 8 bands DFT calculations required approximately 9 hours on 2 nodes with 16 OMP threads each, while the TB calculations required approximately 1 hour on one node with 16 OMP threads. 


\subsection{Current and optical/UV absorption in hBN with excitons}

In this section, we investigate hBN by using the same conditions as before, with the same laser parameters, but now turning on the excitonic (electron-electron) interactions.

First, we calculate the light-induced current, see figure \ref{fig:Abs}e. In comparison with our previous calculations, when interactions are not considered, the current is not damping out and there is a clear current oscillation after the laser pulse. Comparing the Fourier transform of the current without and with interactions, see figures \ref{fig:Abs}d and f respectively, clear differences are appreciable with the TB and KS Hamiltonians showing a very similar behavior. Second, we calculate the absorption spectrum, see figure \ref{fig:Abs}h. The spectrum shows the characteristic exciton peaks below the bandgap at 4.5 eV. The absorption quickly decreases after 5 eV. Interestingly, the TB model and KS Hamiltonian show very similar behavior. 

We also compare the results with those obtained by solving the BSE equations for the TB model. The BSE is based on solving the time-independent Schr\"odinger equation in order to obtain the energies and wavefunctions of excitons. The BSE energies are given by vertical blue and gray lines in figure \ref{fig:Abs}h. We observe that the agreement is excellent for the blue lines. The gray lines correspond to dark excitons that are not excited by the laser pulse. This is in agreement with previous analysis of excitons in hBN \cite{Galvani2016}, in which two of the lowest-energy excitons are dark due to their symmetry. 

The calculations were performed in Intel Xeon CPUs type E5-2630 with 2.40 GHz. The size of the reciprocal space grid is $N=300\times300$ points and the number of terms in the Fourier series expansion is $N_{cut}=20$. The time step is controlled dynamically and varies in the range of 1.2 - 0.3 as during the simulation up to 80 femtoseconds. The 8 bands DFT calculations required approximately 11 hours on 8 nodes with 16 OMP threads each, while the TB calculations required 11 hours on one node with 16 OMP threads. 

These calculations demonstrate the feasibility of the dynamical mean-field approximation to describe excitons, whose energies are the same as those obtained by solving BSE equations, in which the two-particle Hamiltonian is fully accounted. This opens the door to investigate excitonic interactions in real time for pump-probe and nonlinear schemes.

\subsection{ATAS in graphite}

\begin{figure} 
\includegraphics[width=0.49\textwidth]{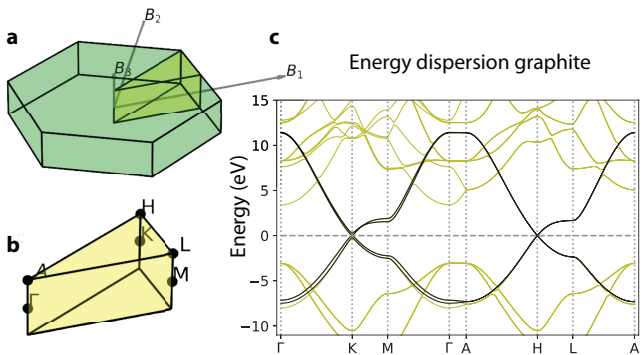}
  \caption{a) First Brillouin zone of graphite and reciprocal lattice vectors. b) Main points of the reciprocal space. c) Energy dispersion along the path $\Gamma - K - M - \Gamma - A - H - L - A$.
  }\label{fig:graphite1}
\end{figure}

We model an attosecond ultrafast spectroscopy experiment in a few-layer graphite. In particular, we model the changes of absorption of a X-ray attosecond pulse that goes through a material interacting with a mid-IR ultrashort pulse, i.e. we calculate the ATAS spectrum. These ultrafast experiments are currently feasible in advanced laser laboratories based on high-harmonic generation sources \cite{Leone2016}.

\begin{figure*}
\includegraphics[width=0.99\textwidth]{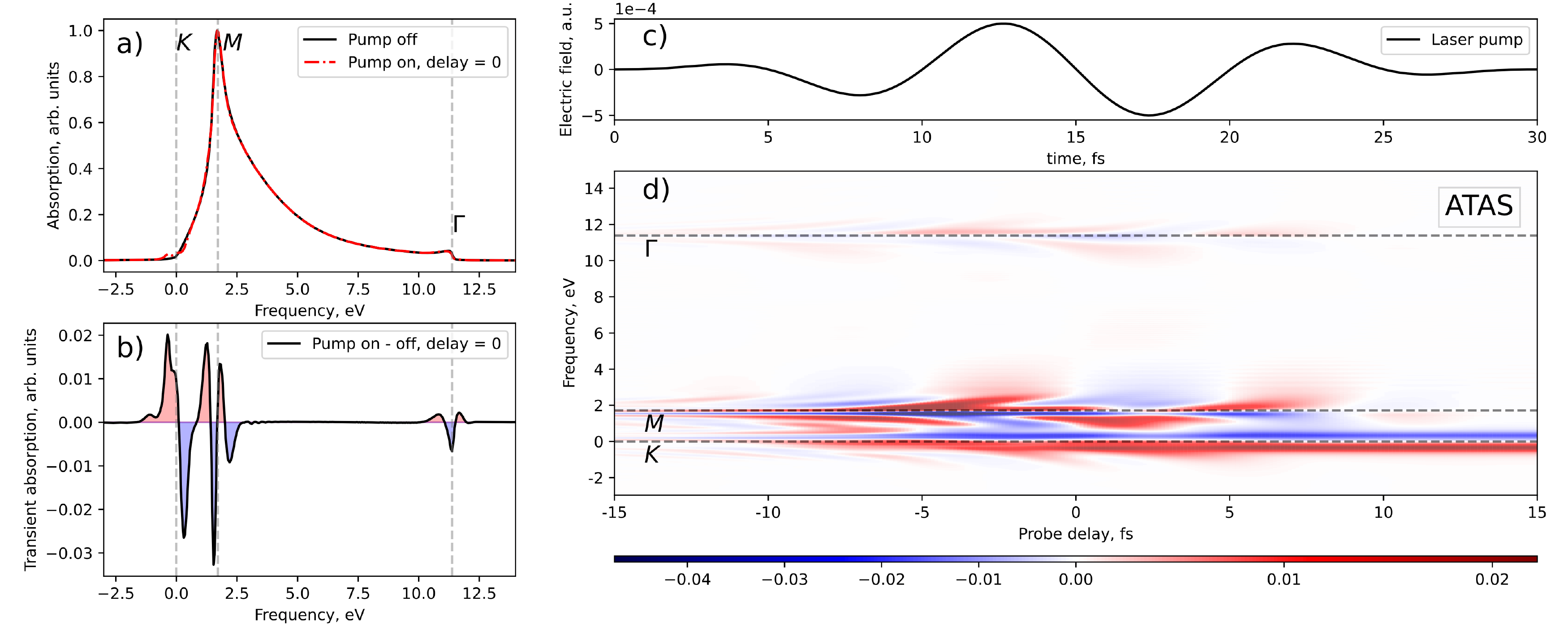}
  \caption{ X-ray attosecond transient absorption in graphite. a) The calculated absorption spectrum with (red line) and without (black line) the IR pulse in the case of zero delay. b) the difference between the two spectra, with and without the IR pulse, that shows changes due to the transient dynamics. The time delay between the two pulses is 0 fs (the maximum points of the pulse envelopes are overlapping). c) IR pulse in time. d) ATAS spectrum. Negative time delays correspond when the attosecond pulse arrives first.
  }\label{fig:graphite2}
\end{figure*}

For the simulations we consider photon energies close to the resonant transitions between the 1s bands and the conduction bands, which are around 296 eV. The energy dispersion for graphite is shown in Fig. \ref{fig:graphite1} and is obtained by using the Quantum Espresso code \cite{QuantumEspresso}. In graphite we have many monolayers of graphene that are separated by a distance of 3.35Å. The unit cell contains 4 atoms of carbons. The first Brillouin zone is shown in Fig. \ref{fig:graphite1}a. After calculating the electronic structure with Quantum Espresso, we create four Wannier orbitals by using the Wannier90 code \cite{Wannier90}, which perfectly reproduce the energies of the four bands close to the Fermi level, see Fig. \ref{fig:graphite1}c. In order to take into account the X-ray excitations, we also include the 1s orbitals of the four carbons in the unit cell, which are taken from electronic calculations of isolated carbon atoms \cite{Bunge1993}. The four core bands are degenerate in energy. Because the 1s orbitals are well-localized, the corresponding energy-dispersion bands are flat. With the total of eight orbitals, we calculate the Berry connections that are used in our electron dynamics code. 

We consider a weak X-ray pulse, intensity 10$^9$ W/cm$^2$, linearly polarized along the $\Gamma - A$ direction, and a pulse duration of 80-as FWHM. The bandwidth of the pulse is large enough to cover the whole energy range of the four bands close to the Fermi level, i.e. the bandwidth is larger than 18 eV. The mid-IR pulse has a wavelength of 3000 nm and a moderate intensity of 10$^{10}$ W/cm$^2$, enough intensity to promote electron carriers around the Fermi level. The laser is linearly polarized along the $\Gamma-K$ direction and it is very short, only 30 fs (around 3 cycles). The core-hole decay of the carbon is dominated by Auger processes. We include a core-hole decay of width $\Gamma_{ch}=0.108$ eV, corresponding to a lifetime of 6.1 fs, the expected lifetime of the 1s vacancies at a carbon atom. The bands are occupied below the Fermi level, i.e. $T=0$ distribution. 

The calculated X-ray absorption spectrum is shown in Fig. \ref{fig:graphite2}a for a 0 fs time delay, i.e. when the peaks of the envelopes of both pulses are overlapping. The zero in the energy scale corresponds to the transition from the 1s band to the Fermi level. We show the calculated absorption without the mid-IR pulse, see the black line. We repeat the simulation, but in the presence of the mid-IR laser pulse and we observe clear changes in the absorption, see the red line. By taking the difference, we obtain the transient absorption spectrum with changes that are mainly localized around van Hove singularities, see Fig. \ref{fig:graphite2}b. This is similar to what has been reported in graphene \cite{Cistaro2021}. We obtain the ATAS spectrum by repeating the same transient absorption calculations for different time delays between the X-ray attosecond and mid-IR pulse, see Fig. \ref{fig:graphite2}d. For reference, we show the mid-IR pulse in time in Fig. \ref{fig:graphite2}c. The changes around the Fermi level (such as K and H points) are due to the promotion of electrons from the valence to the conduction, which affects the excitation process itself due to Pauli exclusion. The other two features, around 2 eV and 12 eV (where the M,L and $\Gamma$,A points are located respectively), are related to the coherent laser-driven dynamics of electrons in the conduction band promoted from core bands \cite{Cistaro2021}. We should remark here the importance of treating both pump and probe pulses on an equal footing in order to obtain the transient absorption spectrum. With the current technology \cite{Buades2021}, it will be possible to carry out an ultrafast experiment similar to the one modeled in this work. 

The calculations were performed in Intel Xeon Platinum 8160 with 2.10 GHz. The size of the reciprocal space grid is $N=200\times200\times10$ points. The time step is 2.4 as during the simulation up to 80 femtoseconds. The calculations for each time delay took 12 hours using 4 CPUs with 48 OMP threads.

In summary, these calculations demonstrate the possibility to use Wannier orbitals to compute real-time simulations as well as to describe ultrafast spectroscopy studies. 

\section{\label{sec:conclusions}Conclusions}

We have presented a theoretical framework and numerical implementation to simulate the out-of-equilibrium electron dynamics that arises when a condensed-matter system is irradiated by ultrashort laser pulses. The approach is based on real-time simulations in the first Brillouin zone by imposing periodic boundary conditions. We solve the equation of motion of the reduced one-electron density matrix of the system and calculate observable expectation values in the time domain, such as light-induced current and polarization in the medium. The approach is suitable to account for: i) light-matter interactions for different laser pulses, ii) wide range of photon energies within the mid-IR to the soft-X-ray regime, and iii) excitonic interactions within the dynamical mean-field approximation, and iv) to work with any localized Bloch basis. Furthermore, the approach can work with the electronic structure previously calculated with a DFT code. We show the robustness of the code to work with local orbitals and Wannier basis from CRYSTAL and Wannier90 calculations. As a relevant example, we show the calculations of the current and optical/UV spectrum for hBN with and without excitons via real-time simulations. Also, we model a realistic absorption pump-probe experiment with a mid-IR laser and an attosecond X-ray pulse in graphite.

Due to the flexibility of our approach to use Bloch states on a local orbital basis previously calculated from DFT calculations, we envision the possibility to describe the light-induced response in functional materials of current interest, such as the photogalvanic effect and the HHG emission, and also to describe novel ultrafast spectroscopy studies.

\begin{appendices}

\section{Absorption and conductivity in linear response}
\subsection{Linear response using time-dependent perturbation theory}
The absorption calculated using equation (\ref{Absorption}) can be compared with the formula of the absorption obtained in first-order linear response approximation as long as we are in the weak intensity regime. We consider the absorption of a photon of $\hbar \omega$ energy via the energetic transition $\Gamma_{f \rightarrow i}$, being $\Gamma_{f \rightarrow i}$ the number of transitions per unit time per unit volume, normalized to a linearly polarized incident flux $I(\omega)=2 c\epsilon_0|\boldsymbol{\varepsilon}(\omega)|^2$ as 
\begin{equation}
    \alpha(\omega)=\sum_{f}\frac{\Gamma_{i \rightarrow f}(\omega)\hbar \omega}{2 c\epsilon_0|\boldsymbol{\varepsilon}(\omega)|^2}.
\end{equation}
Here $\boldsymbol{\varepsilon}(\omega)$ refers to the Fourier transform of the electric field with positive frequencies. All transition processes from the initial state are considered here. The transition rate is now worked out by using time-dependent perturbation theory (similarly to using Fermi's Golden Rule \cite{Cohen}):
\begin{eqnarray}
\Gamma_{f \rightarrow i} = \frac{1}{ V \hbar^2} \frac{d}{dt}\Big| \int_{0}^{t} dt' \braket{f|\Big(|e| \boldsymbol{\varepsilon}(t')\cdot \hat{\bf r} \Big)|i} e^{-{i\over \hbar}(\epsilon_i-\epsilon_f)t'}\Big|^2  \nonumber \\
\end{eqnarray}
where $\epsilon_f$ and $\epsilon_i$ correspond to the unperturbed energies of the final and initial eigenstates, respectively. Inserting the Fourier transform of the electric field, defining $\epsilon_f = \hbar\omega_f $ and $\epsilon_i = \hbar\omega_i$, and performing the integration in time,
\begin{eqnarray}
\Gamma_{f \rightarrow i} =  \frac{|e|^2}{ V \hbar^2} \frac{d}{dt}\Bigg| \int_{-\infty}^\infty d\omega \, \boldsymbol{\varepsilon}(\omega)\cdot \braket{f|\hat{\bf r} |i} \Big[\frac{e^{i(\omega_f-\omega_i-\omega)t}-1}{\omega_f-\omega_i-\omega}\Big]\Bigg|^2 \nonumber \\
\end{eqnarray}
Assuming that the main contribution of the integral is for frequencies very close to the transition resonance, then it is a good approximation to reduce the square modulus to the integrand \cite{Cohen}, i.e.
\begin{eqnarray}
\Gamma_{f \rightarrow i} =  \frac{|e|^2}{ V \hbar^2} \int_{-\infty}^\infty d\omega  |\boldsymbol{\varepsilon}(\omega)\cdot \braket{f|\hat{\bf r} |i}|^2 \frac{d}{dt} \Bigg| \frac{e^{i(\omega_f-\omega_i-\omega)t}-1}{\omega_f-\omega_i-\omega} \Bigg|^2. \nonumber \\
\end{eqnarray}
In the limit $t\longrightarrow \infty$ the above oscillating function tends to $2\pi \hbar t \delta(\hbar \omega-[\epsilon_f-\epsilon_i])$. After the time derivative and considering only the positive frequencies that satisfy the energy conservation
\begin{eqnarray}
\Gamma_{f \rightarrow i} & = & \int_0^\infty d\omega \, \Gamma_{f \rightarrow i}(\omega)  \nonumber \\
& = & \int_0^\infty d\omega \frac{2 \pi \hbar|e|^2}{ V \hbar^2} |\boldsymbol{\varepsilon}(\omega) \cdot \braket{f|\hat{\bf r} |i}|^2 \delta(\hbar \omega-[\epsilon_f-\epsilon_i]). \nonumber \\
\end{eqnarray}
Hence, the absorption depending on the frequency is cast as
\begin{eqnarray}
\alpha(\omega)&=\frac{\pi |e|^2 \omega}{ c \epsilon_0 V }  \sum_{f}   | {\bf u} \cdot \braket{f|\hat{\bf r} |i}|^2 \, \delta(\hbar \omega-[\epsilon_f-\epsilon_i]), \, \label{absorptionlr}
\end{eqnarray}
where ${\bf u}$ is a unitary vector indicating the polarization direction of the incident field. In linear response approximation the frequency-dependent polarization is linked to the conductivity as ${\bf P}(\omega)\vert_k = -i\omega^{-1}{\bf J}(\omega)\vert_k= -i\omega^{-1} \sigma_{kj} (\omega){\boldsymbol \varepsilon}(\omega)\vert_j$, being ${\sigma}_{kj}$ the two-rank conductivity tensor. In linear response, the absorption is related with the first-order susceptibility $\chi$ as
\begin{eqnarray}
\alpha(\omega) = {\omega \over c \epsilon_0} {\rm Im}[\chi_{\bf u}]
\end{eqnarray}
where ${\bf P}(\omega)\vert_k = \chi_{kj} (\omega){\boldsymbol \varepsilon}(\omega)\vert_j$ and ${\bf u}$ refers to the polarization direction of the field. The conductivity and the susceptibility are related then by $\sigma_{kj} (\omega) = i\, \omega\, \chi_{kj} (\omega)$.  Using this relation and the absorption formula given by Eq. (\ref{absorptionlr}), the real part of the optical conductivity is
\begin{eqnarray}
 \text{Re}[\sigma (\omega)]&=& c \epsilon_0 \, \alpha(\omega) \nonumber\\
 &=& \frac{ \pi |e|^2 \omega}{  V }  \sum_{f}   | {\bf u}\cdot \braket{f|\hat{\bf r} |i}|^2 \delta(\hbar \omega-[\epsilon_f-\epsilon_i]). \nonumber \\
\end{eqnarray}

\subsection{Conductivity: single particle approximation and excitons}
The formula above must be particularized for a set of final excited states. Considering electron-hole correlations, exciton states are written as $\ket{X_N}=\Sigma_{vc {\bf k}}A_{cv}^{(N)}({\bf k})\hat{c}_{c}^{\dagger}\hat{c}_{v}\ket{GS}$, where one uses a basis of single-particle excitations states from valence (v) to conduction bands (c). The wavefunction $A_{cv}^{(N)}(\rm k)$ is typically found by solving the Bethe-Salpeter equation \cite{Rohlfing1998}. $N$ is the number of exciton states. In our case, this number is related to the number of points in the reciprocal space. The optical conductivity reads then \cite{Ridolfi2018}
\begin{equation} \label{kuboexciton}
\text{Re}[\sigma(\omega)]= \frac{ \pi |e|^2 \omega}{  V }  \sum_{N}   |{\bf u} \cdot \braket{X_N|\hat{\bf r} |GS}|^2 \delta(\hbar \omega-E_N),
\end{equation}
The matrix elements can be computed using the single-particle states as $\braket{X_N|\hat{\bf r} |GS}=\Sigma_{cv {\rm k}}A_{cv}^{(N)}({\bf k}) \braket{c {\bf k}|\hat{\bf r}_1 |v {\bf k}}$. In the absence of interactions, none of the electron-hole pairs are correlated and optical conductivity formula in linear response is reduced to the well-known Kubo-Greenwood formula
\begin{equation}
\text{Re}[\sigma(\omega)]= \frac{ \pi |e|^2 \omega}{  V }  \sum_{cv {\bf k}}|{\bf u} \cdot \braket{c {\bf k}|\hat{\bf r}_1 | v {\bf k}}|^2 \delta(\hbar \omega-[{\epsilon}_f({\bf k})-{\epsilon}_i({\bf k})]).
\end{equation}



\end{appendices}

\begin{acknowledgments}
G. Cistaro, M. Malakhov, and A. Pic\'{o}n acknowledge Comunidad de Madrid through TALENTO grant ref. 2017-T1/IND-5432, grant ref. RTI2018-097355-A-I00 (MCIU/AEI/FEDER, UE), and computer resources and assistance provided by Centro de Computaci\'on Cient\'ifica de la Universidad Aut\'onoma de Madrid (FI-2021-1-0032), Instituto de Biocomputación y Física de Sistemas Complejos de la Universidad de Zaragoza (FI-2020-3-0008), and Barcelona Supercomputing Center (FI-2020-1-0005, FI-2021-2-0023, FI-2021-3-0019). J. J. Palacios J. J. Esteve-Paredes, and A. J. Uría-Álvarez acknowledge funding from Grant No. PID2019-109539GB-C43 (MCIU/AEI/FEDER, UE), the María de Maeztu Program for Units of Excellence in R\&D (Grant No. CEX2018-000805-M), the Comunidad Autónoma de Madrid through the Nanomag COST-CM Program (Grant No. S2018/NMT-4321), and the Generalitat Valenciana through Programa Prometeo/2021/01. F. Martín acknowledges the MICIN project PID2019-105458RB-I00, the ``Severo Ochoa” Programme for Centres of Excellence in R\&D (SEV-2016-0686), and the ``María de Maeztu” Programme for Units of Excellence in R\&D (CEX2018- 000805-M).  R.E.F.S. acknowledges support from the fellowship LCF/BQ/PR21/11840008 from “La Caixa” Foundation (ID 100010434).
\end{acknowledgments}




\nocite{*}


\end{document}